\documentclass[11pt]{article}

\usepackage{amssymb, amsmath, amsthm, graphicx}
\usepackage[left=1in,top=1in,right=1in]{geometry}

\usepackage{subfigure}
\usepackage{enumerate}

      \theoremstyle{plain}
      \newtheorem{theorem}{Theorem}[section]
      \newtheorem{lemma}[theorem]{Lemma}
      \newtheorem{corollary}[theorem]{Corollary}
      \newtheorem{proposition}[theorem]{Proposition}

      \theoremstyle{definition}

      \theoremstyle{remark}

\title{On Sets of Lines Not Supporting Trees}

\author{
        Radoslav Fulek\thanks{Ecole Polytechnique F\'ed\'erale de Lausanne, {\tt radoslav.fulek@epfl.ch}.
        The author gratefully acknowledge support from the Swiss National
Science Foundation Grant No. 200021-125287/1}
        \hspace{14mm}
        Daniel Neuwirth\thanks{Wilhelm-Schickard-Institut f\"{u}r Informatik – Universit\"{a}t T\"ubingen, Germany {\tt
        neuwirth@informatik.uni-tuebingen.de}}
}

\begin{document}
\date{}

\maketitle

\begin{abstract}
We study the following problem introduced by Dujmovic et al in \cite{dujmovic}.
Given a tree $T = (V,E)$, on $n$ vertices, a set of $n$ lines  $\mathcal{L}$ in the plane  and
a bijection $\iota: V \rightarrow \mathcal{L}$, we are asked to find a crossing-free straight-line embedding of $T$ so that
$v\in \iota(v)$, for all $v\in V$. We say that a set of $n$ lines $\mathcal{L}$  is universal for trees if for
any tree $T$ and any bijection $\iota$ there exists such an embedding.
We prove that any sufficiently big set of lines is not universal for trees, which solves an open problem
asked by Dujmovic et al.
\end{abstract}


\section{Introduction}

Let $G=(V,E)$ denote a simple graph on $n$ vertices.
Throughout this article by a \emph{geometric graph} we understand a representation of the graph $G$ in the plane in which the vertices are represented
by $n$ distinct points and the edges are drawn as straight-line segments.
A geometric graph is crossing free if the relative interior of every edge is disjoint from the rest of the graph,
i.e. if it is an \emph{embedding}.  We do not distinguish between an abstract graph and a geometric graph, and we use  ``vertex'' and ``edge''  in both contexts.

We say that a set $P$ of points in the plane is $n$-universal, if any planar graph on $n$ vertices admits a crossing free representation as a geometric graph
in which the vertices are represented by the points of $P$.

The problem of finding a smallest $n$-universal point set stimulated a significant amount of research, see e.g.~\cite{brandenburg,chrobak,deFraysseix,everett,kurowski,schnyder}.
Brandenburg~\cite{brandenburg} showed  that a set of $\frac 89 n^2$ points forming the
 $\frac 43 n \times \frac 23 n$ grid is $n$-universal, which is also the best known upper bound on the size of an $n$-universal point set. On the other hand, Chrobak and Karloff proved in~\cite{chrobak}  that for sufficiently high $n$, an $n$-universal point set is of size at least $1.089n$, which was later improved to $1.235n$ by Kurowski~\cite{kurowski}.

We treat an analogous problem in which we consider line sets instead of point sets. However, in case we define a set of lines $\mathcal{L}$ to be $n$-universal, if any planar graph on $n$ vertices admits a representation as a geometric graph such that
each of its vertices lies on a unique line in $\mathcal{L}$, any set of $n$ lines in the plane is  $n$-universal~\cite{dujmovic2}. This is also a consequence of the main result in~\cite{janos}.
Hence, in case of line sets we consider a stronger definition of universality, the one that was introduced in~\cite{dujmovic}.

Thus, a set of lines $\mathcal{L}$ \emph{supports} a planar graph $G=(V,E)$, if for any  bijection $\iota:V \rightarrow \mathcal{L}$
there exists a crossing-free representation of $G$ as a geometric graph such that $v\in \iota(v)$.
We say that a set of lines $\mathcal{L}$ is  \emph{universal (for trees)} if it supports any planar graph (tree) on $|\mathcal{L}|$ vertices.
Hence, contrary to the universality in case of point sets, we require that every line in the set accommodate a vertex.

It was shown in~\cite{dujmovic} that there exists a line set which is not universal, and that no sufficiently big set of concurrent lines is universal.
Later Dujmovi\'c and Langerman in~\cite{dujmovic2} improved this result by showing that no sufficiently big set of lines is universal.
The main purpose of this note is
a strengthening of their result by showing that no sufficiently big set of lines is universal even for trees:

\begin{theorem}
\label{thm:concurrentLines}
There exists a constant $n_0$ such that no set of lines of size $n,  n\ge n_0$, is  universal for trees.
\end{theorem}

On the other hand it is known~\cite{dujmovic} that any set of lines is universal for lobsters, i.e.  trees containing a path
reachable from every vertex by a path of length at most two.

The article is organized as follows. In Section 2, we show some geometric Ramsey-type results and one geometric lemma that allow us to concentrate only
on certain regular line arrangements in the proof of the main result, which is deferred to Section 3. We conclude in Section 4 with some remarks and
possible extension of our result.


\section{Preliminaries}

Let $\mathcal{L}$ denote a set of lines $\{l_1,\ldots,l_n\}$ none of which is vertical such that the lines in $\mathcal{L}$ are indexed increasingly according to their slopes, i.e. for the slopes $s(l_i)$ of lines in
$\mathcal{L}$ we have $s(l_i)<s(l_j)$, if $i<j$. We assume that no three lines in $\mathcal{L}$ meet in a point and no two lines in $\mathcal{L}$ are parallel. By the \emph{angle} $a(l)$ of the line $l$ we understand $\arctan(s(l))$.
Let $D(\mathcal{L})$ denote the set of points dual to $\mathcal{L}$ in the following point-line duality $(a,b) \leftrightarrow y=ax-b$.

\begin{figure}
  \centering
  \subfigure[]{
  \includegraphics[scale=0.25]{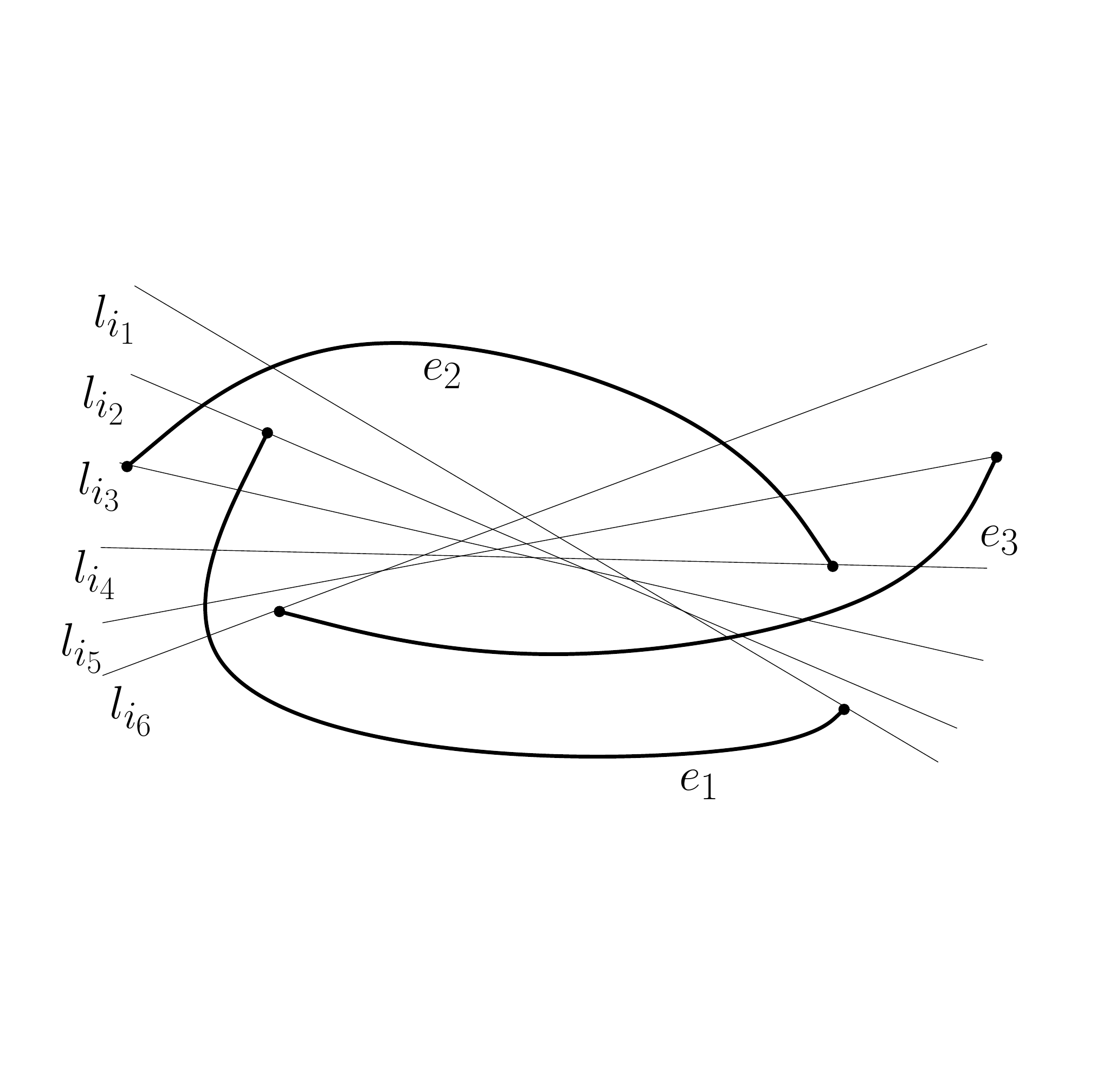}
  \label{fig:unstretch}
  }  \hspace{1cm} \subfigure[]{
  \includegraphics[scale=0.35]{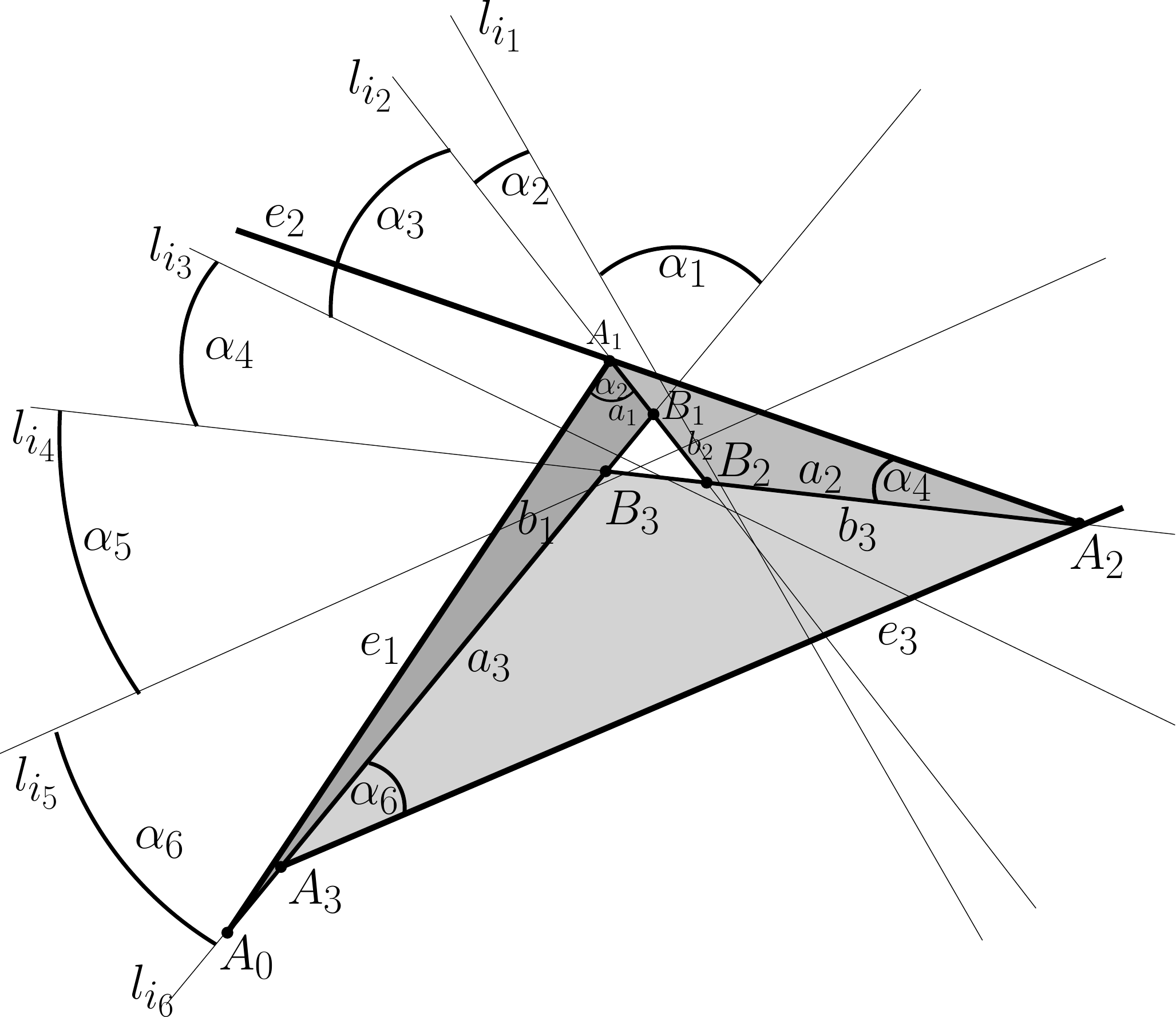}
  \label{fig:unstretch2}
  }
  \caption{(a) Unstretchable configuration of three edges connecting pairs of lines of $\mathcal{L}'$ , (b) Configuration of geometric objects in the proof Lemma~\ref{lemma:unstretch}.   (Due to a better readability the figure is slightly misleading.  In particular, the edge $e_1$
should be almost parallel to $l_{i_1}$ and $l_{i_2}$.)}

\end{figure}

In the proof of our result we focus only on a regular subset of lines $\mathcal{L}$ called {\em cap} (resp. {\em cup}) into which we map vertices of a subtree of our given tree.
We say that the set of lines $\mathcal{L}$ forms a {\em cap} (resp. {\em cup}) (lines in Fig.~\ref{fig:unstretch} form a cap), if the intersections of $l_i$ with the lines $l_1,\ldots l_{i-1},l_{i+1},\ldots l_n$, for $i=1,\ldots, n$,  appears along $l_i$ from right to left (resp.  left to right) in that order.
The notion of {\em cap} (resp. {\em cup}) is often used in the literature in the context of point sets, where it stands for a set of points in the plane in a strictly convex position such that
there exists a concave (resp. convex) function passing through all the points the set.
By the point-line duality, $\mathcal{L}$ forms a  cap or cup if and only if $D(\mathcal{L})$ forms a  cup or  cap, respectively. Hence, by the famous Erd\H{o}s-Szekeres Theorem, $\mathcal{L}$ contains a subset $\mathcal{L}'$ of lines which forms  a cap or cup
of size $\Omega(\log n)$.

Let $H=(V,E)$ denote the complete three-uniform hypergraph with the vertex set $V=\{1,\ldots,n\}$.
We call a subset $P\subseteq E$ a \emph{path} of length $k$, if $P$ is of the form $\{\{i_{j},i_{j+1},i_{j+2}\} | \ 1\leq j\leq k-2 \}$ for
$1\leq i_1<i_2<\ldots<i_k\leq n$.

The next lemma is well-known as its proof follows easily from the proof of Erd\H{o}s-Szekeres Theorem. The proof can be
found e.g. in~\cite{zuk}.

\begin{lemma}
\label{lemma:3reg}
If we two-color edges of $H$, then $H$ contains a path of length $\Omega(\log n)$, all of whose edges have the same color (i.e.
a monochromatic path).
\end{lemma}

As a simple corollary of Lemma \ref{lemma:3reg} we get a result which allows us to select in $\mathcal{L}$ a subset of lines $\mathcal{L'}$ so that the angle difference between two consecutive lines in $\mathcal{L'}$ is non-decreasing (or non-increasing).

\begin{corollary}
\label{cor:3reg}
For any set of lines $\mathcal{L}$ there exists a subset $\mathcal{L}'=\{l_{i_1},\ldots,l_{i_k}\}$, $i_1<i_2\ldots <i_k$, of $\mathcal{L}$ of size $\Omega(\log n)$ so that
the following sequence of angles is non-decreasing (or non-increasing): \\
$(a(l_{i_{2}})-a(l_{i_{1}}),a(l_{i_{3}})-a(l_{i_{2}}), \ldots , a(l_{i_{k}})-a(l_{i_{k-1}}))$.
\end{corollary}

\begin{proof}
Let $H=(\mathcal{L},E)$ denote the complete three-uniform hypergraph. We color the edge $\{l_{i_1},l_{i_2},l_{i_3}\}$ by red
if $a(l_{i_{3}})-a(l_{i_{2}})< a(l_{i_{2}})-a(l_{i_{1}})$ and by blue if $a(l_{i_{3}})-a(l_{i_{2}})\geq a(l_{i_{2}})-a(l_{i_{1}})$.
By applying Lemma \ref{lemma:3reg} on $H$ we get a monochromatic path $P$ of length $\Omega(\log n)$. It is easy to check that the vertex
set of $P$ is the required set of lines $\mathcal{L}'$.
\end{proof}

In fact, in the sequel we use the following statement, which is easy to obtain from
Corollary \ref{cor:3reg}.

\begin{corollary}
\label{cor:increasing}
For any set of lines $\mathcal{L}$ there exists a subset $\mathcal{L}'=\{l_{i_1},\ldots,l_{i_k}\}$; $i_1<i_2\ldots <i_k$; $a(l_{i_1})<\ldots < a(l_{i_k})$, of $\mathcal{L}$ of size $\Omega(\log\log n)$ so that
$|a(l_{i_k})-a(l_{i_1})|<\frac \pi2$ and for each $2\leq j\leq k-1$ the following holds: \\
$a(l_{i_{j+1}})-a(l_{i_{j}})\geq a(l_{i_{j}})-a(l_{i_{1}})$ $($or $a(l_{i_{j}})-a(l_{i_{j-1}})\geq a(l_{i_{k}})-a(l_{i_{j}}))$.
\end{corollary}

\begin{proof}
It is easy to verify that given a subset $\mathcal{L}''=\{l_{i_1},\ldots,l_{i_l}\}$ of $\mathcal{L}$ from Corollary \ref{cor:3reg},
we can take as $\mathcal{L}'$ the following subset of  $\mathcal{L}''$: $\{l_{i_{2^0}},l_{i_{2^1}},\ldots,l_{i_{2^{\lfloor log l \rfloor-1}}}\}$.
\end{proof}

In the rather long and technical proof of our main result we direct our effort towards finding an unstretchable
configuration of three edges connecting three pairs of lines of $\mathcal{L}'$ from Corollary~\ref{cor:increasing}.
Ref. to Fig.~\ref{fig:unstretch}. The desired configuration of three edges $e_1,e_2$ and $e_3$  has the following properties.
Let $l_{i_1},\ldots, l_{i_6}\in \mathcal{L}'$ (as in Corollary~\ref{cor:increasing}).
\begin{enumerate}[(i)]
\item
The edge $e_j$ joins $l_{i_{2j}}$ with $l_{i_{2j-1}}$ so that $e_j$ passes below the intersection point of
$l_{i_{2j}}$ and $l_{i_{2j-1}}$, if $j\in\{1,3\}$ and above otherwise;
\item
The edge $e_j$  is disjoint from the convex hull of the intersection points of $l_{i_1},\ldots, l_{i_6}$; and
\item
The endpoint of $e_j$ on $l_{i_{2j}}$ belongs to the line segment between the intersection point of
$l_{i_{2j}}$ and $l_{i_{2j-1}}$, and the intersection point of $l_{i_{2j}}$ and $e_{{j+1} \mod 3}$\footnote{We represent modulo class 0 by 3.}.
\end{enumerate}

Note that properties (i)-(iii) implies that $e_1,e_2$ and $e_3$ do not cross each other.

\begin{lemma}
\label{lemma:unstretch}
The edges $e_1,e_2$ and $e_3$ cannot be drawn as (non trivial) straight-line segments.
\end{lemma}
\begin{proof}
For the sake of contradiction we assume the contrary. Ref. to Fig~\ref{fig:unstretch2}.
Given $e_1,e_2$ and $e_3$ satisfying properties (i)-(iii) from above we slightly rotate $e_j$ around its endpoint on $l_{i_{2j}}$
so that it becomes parallel to  $l_{i_{2j-1}}$, and we shift $e_2$ and $e_1$ so that they touch $e_3$ and $e_2$, resp. This operation causes $e_j$ to intersect $l_{i_{2j-1}}$ at infinity, and hence, from now on $e_j$
is represented as a ray (i.e. half-line) emanating a point on $l_{i_{2j}}$ rather than just a line segment. Moreover, after the rotation and shifting $e_1,e_2$ and $e_3$ still satisfy
properties (i)-(iii) (if we allow crossings at infinity), and thus, they do not cross each other.

Let $A_j$ denote the intersection  of $e_j$ and $l_{i_{2j}}$. Let $B_j$ denote the intersection of  $l_{i_{2j}}$ and $l_{i_{2(j+1 \mod 3)}}$.
Let $a_j, b_j$ and $r_j$ denote $|A_jB_j|, |B_jA_{j-1}|$ and $|B_jB_{j+1 \mod 3}|$, resp. Let $\alpha_j$ for $j>1$ denote the size of the smaller angle between
$l_{i_{j-1}}$ and $l_{i_j}$. Let $\alpha_1=\pi-\sum_{j=2}^{6}\alpha_j$.

We have the following conditions:
{ $b_1-r_3 > a_3$} (1),
{ $\frac{\sin \alpha_2}{\sin \alpha_1}a_1=b_1$}  (2),
{ $b_2-r_1 = a_1$} (3),
{ $\frac{\sin \alpha_4}{\sin \alpha_3}a_2=b_2$} (4),
{ $b_3-r_2 = a_2$} (5),
{ $\frac{\sin \alpha_6}{\sin \alpha_5}a_3=b_3$} (6). \\
\bigskip

\begin{tabular}{lcr}
(5)+(6) $\Rightarrow$ & { $\frac{\sin \alpha_6}{\sin \alpha_5}a_3-r_2= a_2$} & (7)\\
(7)+(4) $\Rightarrow$&{ $\frac{\sin \alpha_4}{\sin \alpha_3}\left(\frac{\sin \alpha_6}{\sin \alpha_5}a_3-r_2\right)= b_2$} & (8) \\
(8)+(3) $\Rightarrow$&{ $\frac{\sin \alpha_4}{\sin \alpha_3}\left(\frac{\sin \alpha_6}{\sin \alpha_5}a_3-r_2\right)-r_1= a_1$} & (9) \\
(9)+(2) $\Rightarrow$&{ $\frac{\sin \alpha_2}{\sin \alpha_1}\left(\frac{\sin \alpha_4}{\sin \alpha_3}\left(\frac{\sin \alpha_6}{\sin \alpha_5}a_3-r_2\right)-r_1\right)=b_1$} & (10) \\
(10)+(1) $\Rightarrow$&{ $\frac{\sin \alpha_2}{\sin \alpha_1}\frac{\sin \alpha_4}{\sin \alpha_3}\frac{\sin \alpha_6}{\sin \alpha_5}a_3  -  \frac{\sin \alpha_2}{\sin \alpha_1}\frac{\sin \alpha_4}{\sin \alpha_3}r_2-\frac{\sin \alpha_2}{\sin \alpha_1}r_1-r_3 > a_3$} & (11)
\end{tabular}\\

The sequence of angles $\alpha_2,\ldots, \alpha_6$ is either non-increasing or non-decreasing.
Since  $|a(l_{6})-a(l_{1})|<\frac \pi2$, $\sin \alpha_1$ is always the biggest among $\sin \alpha_i$-s.
By equation (11) we arrive at contradiction if $\sin \alpha_6,\ldots ,\sin \alpha_2, \sin \alpha_1$ is
a non-decreasing sequence.

 $$\underbrace{\frac{\sin \alpha_2}{\sin \alpha_1}}_{\leq 1}\underbrace{\frac{\sin \alpha_4}{\sin \alpha_3}}_{\leq 1}\underbrace{\frac{\sin \alpha_6}{\sin \alpha_5}}_{\leq 1}a_3  -  \frac{\sin \alpha_2}{\sin \alpha_1}\frac{\sin \alpha_4}{\sin \alpha_3}r_2-\frac{\sin \alpha_2}{\sin \alpha_1}r_1-r_3 > a_3$$

 Otherwise,  $\sin \alpha_2,\ldots ,\sin \alpha_6, \sin  \alpha_1$ is a non-decreasing sequence and
we arrive at contradiction as well since (11) can be rewritten as follows.
 $$\underbrace{\frac{\sin \alpha_6}{\sin \alpha_1}}_{\leq 1}\underbrace{\frac{\sin \alpha_2}{\sin \alpha_3}}_{\leq 1}\underbrace{\frac{\sin \alpha_4}{\sin \alpha_5}}_{\leq 1}a_3  -  \frac{\sin \alpha_2}{\sin \alpha_1}\frac{\sin \alpha_4}{\sin \alpha_3}r_2-\frac{\sin \alpha_2}{\sin \alpha_1}r_1-r_3 > a_3$$.
\end{proof}

\section{Non-Embeddability on a line set in convex position}

\label{se:proof}

The aim of this section is to prove the following theorem from which it is easy to deduce Theorem~\ref{thm:concurrentLines} by using Corollary~\ref{cor:increasing} and Erd\H{o}s-Szekeres Theorem.

Let $\mathcal{L}=\{l_1,\ldots,l_n\}$, $a(l_{1})<\ldots < a(l_{n})$, denote a set of $n$ lines in the plane no two of which are parallel, and no three of which pass through the same point.

\begin{theorem}
\label{thm:concurrentLines2}
A sufficiently large set of lines $\mathcal{L}$ forming a cap or cup such that $|a(l_{n})-a(l_{1})|<\frac \pi2$ and $a(l_{{j+1}})-a(l_{{j}})\geq a(l_{{j}})-a(l_{{1}})$,
 for $j=2,\ldots, n-1$, $($resp. $a(l_{{j}})-a(l_{{j-1}})\geq a(l_{{n}})-a(l_{{j}})$, for $j=2,\ldots, n-1)$ is not an universal line set for trees, i.e.
there exists a constant $n_0$ such that no such a set of lines on more than $n_0$ vertices is  an universal line set for trees.
\end{theorem}

Let us first prove Theorem~\ref{thm:concurrentLines} given that Theorem~\ref{thm:concurrentLines2} holds.

\begin{proof} [Proof of Theorem \ref{thm:concurrentLines}]
By a standard perturbation argument we can assume that no three lines in $\mathcal{L}$ meet in a point and no two lines in $\mathcal{L}$ are parallel.
Let $\mathcal{L}$ denote a set of lines of size $n=c^{c^{c^{n_1}}}$, where $n_1$ is $n_0$ we get from Theorem~\ref{thm:concurrentLines2}, and $c>0$ is an appropriate constant.
Let $\mathcal{L}'\subseteq \mathcal{L}$ denote a subset of lines of $\mathcal{L}$ forming a cup or cap of size $\Omega (\log n)$.
Let $\mathcal{L}''\subseteq \mathcal{L}'$ denote a subset of lines of  $\mathcal{L}'$ we get from Corollary~\ref{cor:increasing} of size  $n_1=\Omega (\log \log \log n)$.
Let $T'=(V',E')$ denote a tree on $\Omega (\log \log \log n)$ vertices, and let $\iota:V'\rightarrow \mathcal{L}''$ denote a mapping,
such that $T'$ does not have a straight-line embedding with $v\in \iota(v)$ for all $v\in V'$. The existence of $T'$ and $\iota$ is guaranteed by Theorem~\ref{thm:concurrentLines2}.

Taking any tree $T=(V,E)$ on $n$ vertices having $T'$ as its subtree and extending the mapping $\iota$ to a mapping $V\rightarrow \mathcal{L}$ proves the theorem.
\end{proof}

\begin{proof} [Proof of Theorem \ref{thm:concurrentLines2}]
W.l.o.g we assume that $\mathcal{L}$ is a set of lines forming a cap and satisfying the condition of the theorem.
We say that an embedding of a tree $T=(V,E)$ \emph{respects} a bijection $\iota:V \rightarrow \mathcal{L}$, if $v\in \iota(v)$.
In what follows we construct a tree $T=(V,E)$ on $n=n_0$ vertices and a bijective mapping $\iota: V \rightarrow \mathcal{L}$ such that there is no straight-line embedding
of $T$ respecting $\iota$.

\paragraph
{Outline} The proof goes as follows. $T=T(d,\Delta)$ is a complete $\Delta$-ary rooted tree missing one leaf (which is purely a technical condition)
 for a sufficiently high $\Delta$ of a sufficient large depth $d$. Hence, $n=\frac{\Delta^{d+1}-1}{d-1}-1$.
 We partition the set of lines $\mathcal{L}$ into constantly many  color classes $\mathcal{L}_1,\ldots,\mathcal{L}_c$ of equal sizes so that each parts contains consecutive lines with respect to the order according to their slopes. Thus, each $\mathcal{L}_i$ contains a constant fraction of lines of $\mathcal{L}$. The parts $\mathcal{L}_i$ impose a grid-like structure on $\mathcal{L}$.

 Next, we define the mapping $\iota$ about which we show that it does not admit a straight-line embedding of $T$ respecting it. We have, in fact, a lot of freedom in how to choose $\iota$, since we only require
 that for each vertex  sufficiently many descendants are mapped to every $\mathcal{L}_i$.
By a Ramsey-type argument the mapping $\iota$ forces, for sufficiently big $\Delta$ and $d$, in any straight-line embedding respecting it constantly many classes $\mathcal{P}^1, \mathcal{P}^2, \ldots$ of pairwise interior disjoint paths such that the paths in each class are uniform with respect to the mentioned grid-like structure.

 To this end we first select a subtree of $T_{d'}$ whose subpaths of root-leaf paths of length at most $d'$ are uniform with respect to the grid-like structure (Proposition~\ref{prop:regular}).
Second, we introduce a notion of a tubus, which is a geometric object defined by a set of uniform paths in $T_{d'}$ emanating from a single vertex, that can be thought of as a pipe-line that entraps some subpaths of its defining paths  in its interior. Due to the properties of the mapping $\iota$ and the fact that two lines cannot cross in the plane more than once, by increasing $\Delta, d$ and $d'$ we increase the length of almost every tubus (Lemma~\ref{lemma:tubusPolygon}).
Moreover, an internal vertex $v$ on a path entrapped in a tubus has many children mapped by $\iota$ to any $\mathcal{L}_i$, which will imply that a tubus of every color type is
emanating from $v$. By the two previous facts and the impossibility of two lines in the plane to cross more than once, it follows that a tubus has to intersect lines in almost
 every color class $\mathcal{L}_i$ between two consecutive crossings with the same line (Lemma~\ref{lemma:tubusPolygon2}). Hence, a tubus and its defining paths is forced to leave  the convex hull of the intersection points of a big subset of $\mathcal{L}$ (Lemma~\ref{lemma:noA}), and wind
around that convex hull.
This reduces the problem essentially to the case when $\mathcal{L}$ is a set of concurrent lines.

Finally, we argue that we can select three edges from three distinct paths each belonging to different classes $\mathcal{P}^i$ that satisfies the hypothesis of Lemma~\ref{lemma:unstretch}.
 \\


We proceed to the detailed description of the above strategy starting with the construction of the mapping $\iota$.
Ref. to Fig.~\ref{fig:regions}.
We partition the lines in $\mathcal{L}$ into $c$ (which is a constant specified later) sets $\mathcal{L}_1,\ldots \mathcal{L}_c$ of equal size (we assume $c \mid n$), such that
$\mathcal{L}_{c'}=\{l_{(c'-1)\frac nc+1},\ldots l_{c'\frac nc}\}$.
Furthermore, we partition the union of lines without their intersection points $\bigcup_{i=1}^{n}l_i\setminus (\bigcup_{i\not=j} (l_i \cap l_j))$ into the regions $R_{a,b}$, $a\leq b$ defined as follows.

Let $P_{i,j}$ denote the $j$-th leftmost intersection point on $l_i$. We define $P_{i,0}$ and $P_{i,n}$, to be the point at $-\infty$ and $\infty$, resp., on $l_i$.
We define on each line $l_i$  open segments
$l_{i,c'}=P_{i,(c'-1)\frac nc}P_{i,c'\frac nc}\setminus \{P_{i,(c'-1)\frac nc},P_{i,c'\frac nc} \}$, $c'=1,\ldots, c$. We set $R_{a,b}=\bigcup_{x=(b-1)\frac nc +1}^{b\frac nc}l_{x,a} \cup
\bigcup_{x=(a-1)\frac nc +1}^{a\frac nc}l_{x,b}$, $a,b=1,\ldots, c$.
The regions $R_{a,b}$ form a grid-like structure on $\mathcal{L}$.

\begin{figure}
  \centering
  \subfigure[]{
  \includegraphics[scale=0.3]{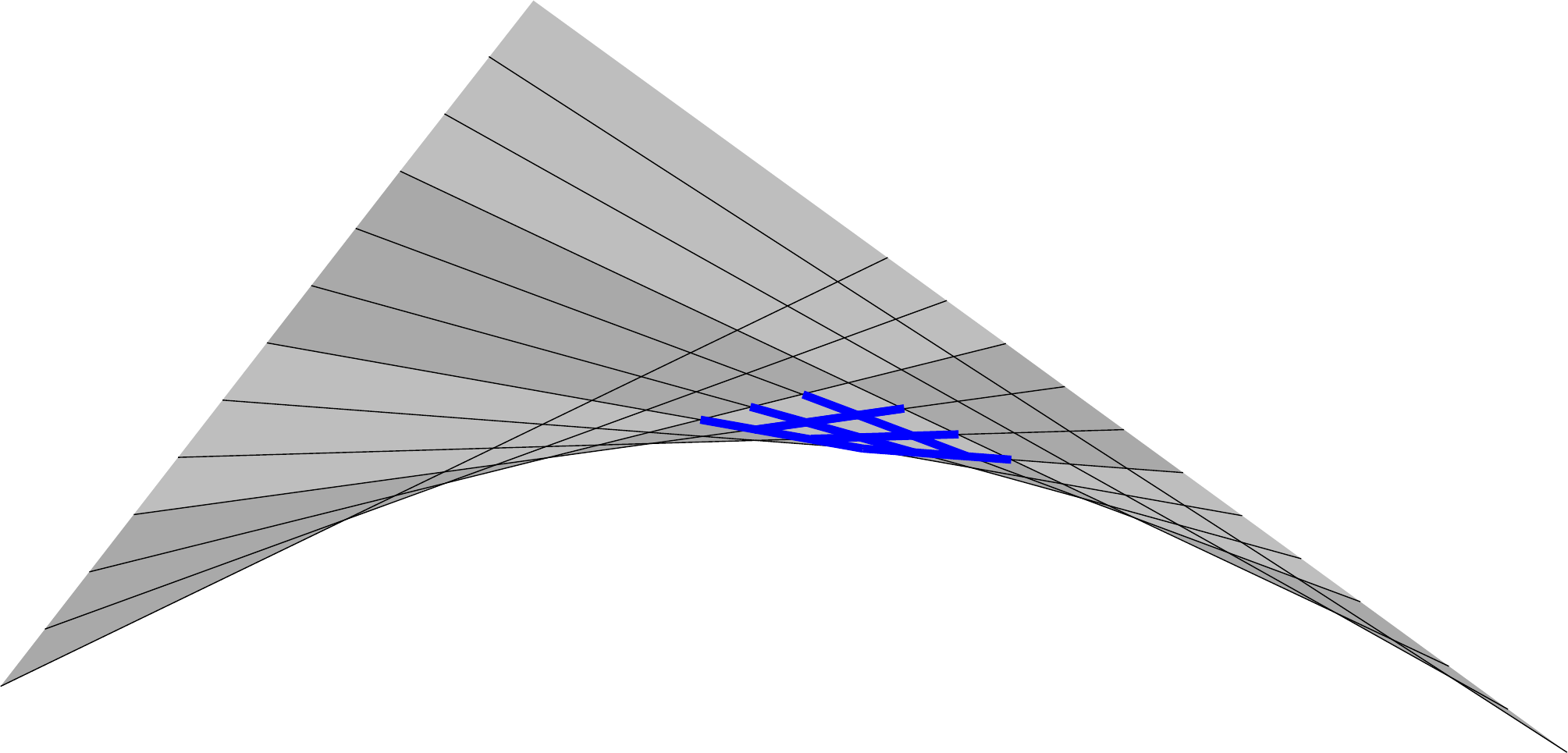}
  \label{fig:regions}
  }  \hspace{1cm} \subfigure[]{
  \includegraphics[scale=0.3]{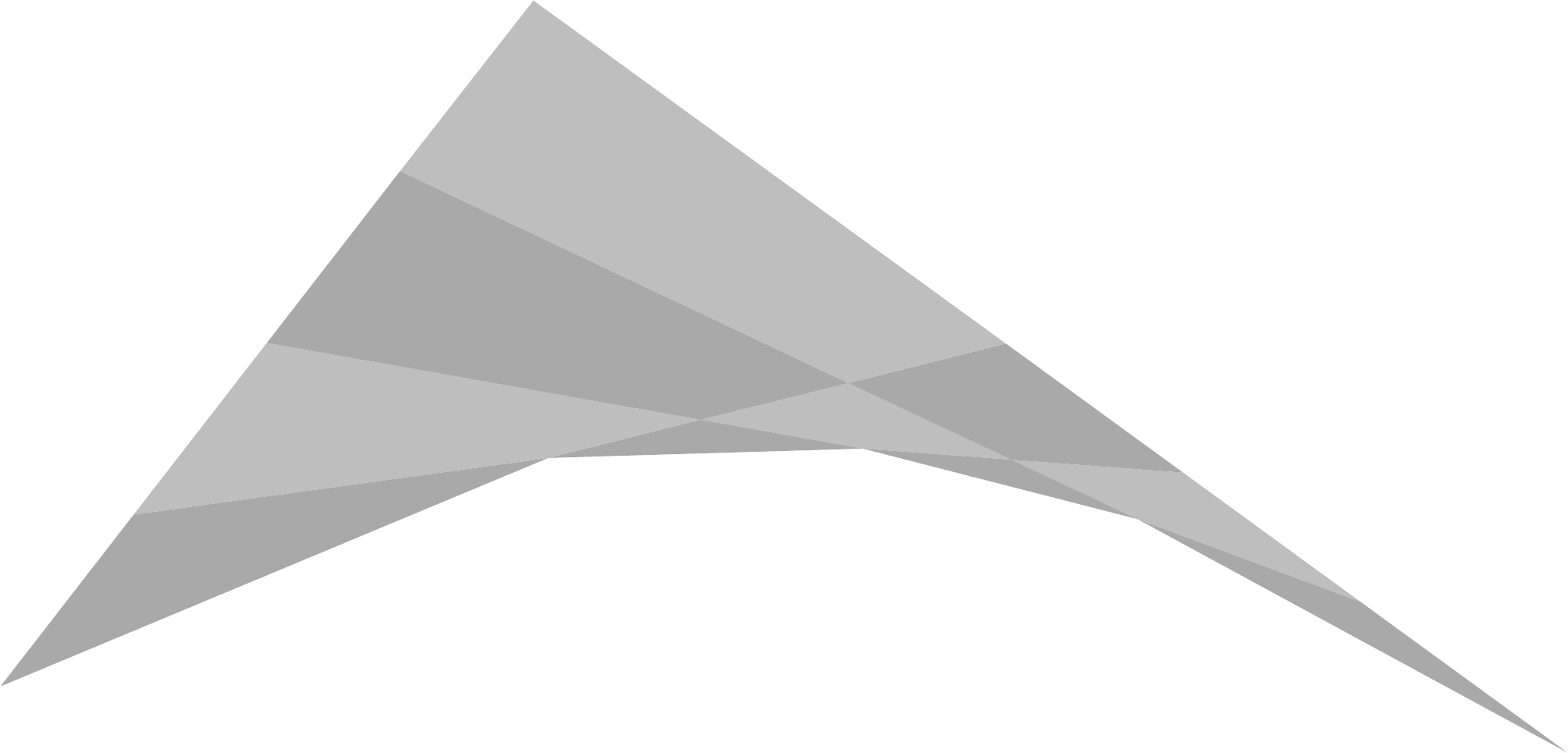}
  \label{fig:regions2}
  }
  \caption{(a) The division into the regions $R_{a,b}$ for $n=12$ and $c=4$ with the highlighted region $R_{2,3}$, (b) Regions $\overline{R_{a,b}}$, if $n=12$ and $c=4$.}

\end{figure}

We define a mapping $\iota$ so that the root is mapped to an arbitrary line in $\mathcal{L}$ (which accounts for the one missing leaf of $T$), and $\frac {\Delta}{c}$ children of every vertex are mapped
arbitrarily to the lines in $\mathcal{L}_{c'}$, for each $c'=1,\ldots ,c$ (we assume $c \mid \Delta$). Thus, one can think of $\iota$ as of a ``typical bijection''
 between $V(T)$ and $\mathcal{L}$ picked uniformly at random.

We orient each edge of $T$ away from the root. Henceforth, the edges of $T$  are directed.
By an {\em (oriented) path} $P$ starting at $v_1$ of length $m$ we understand an ordered $m$-tuple of the vertices $v_1\ldots v_m$, $v_i\in V(T)$, for $i=1,\ldots, m$, such that $\overrightarrow{v_iv_{i+1}}\in E(T)$. We say $v_i\in P$, $1\leq i\leq m$. We call $v_iv_{i+1}$ the $i$-th edge of $P$. By a {\em subpath} $P'$ of $P$ we understand $v_i\ldots v_j$, for some $1\leq i \leq j \leq m$. We say $P'\subseteq P$. By a subpath $P'$ of $P$ in the {\em topological sense} we understand a subcurve of the curve that $P$
corresponds to in our embedding of $T$. We say that two paths are internally disjoint if they do not share a vertex except possibly the vertex they both start at.

For sake of contradiction let us fix a straight-line embedding of $T$ respecting $\iota$.
Let $d'$, $0< d'\leq d$, denote a natural number (we recall that $d$ is the depth of $T$). We select a subtree $T_{d'}$ of $T$ with the same root, which is in some sense regular with respect to our fixed embedding of $T$:

\begin{figure}[htp]
  \centering
   \subfigure[]{\includegraphics[scale=0.3]{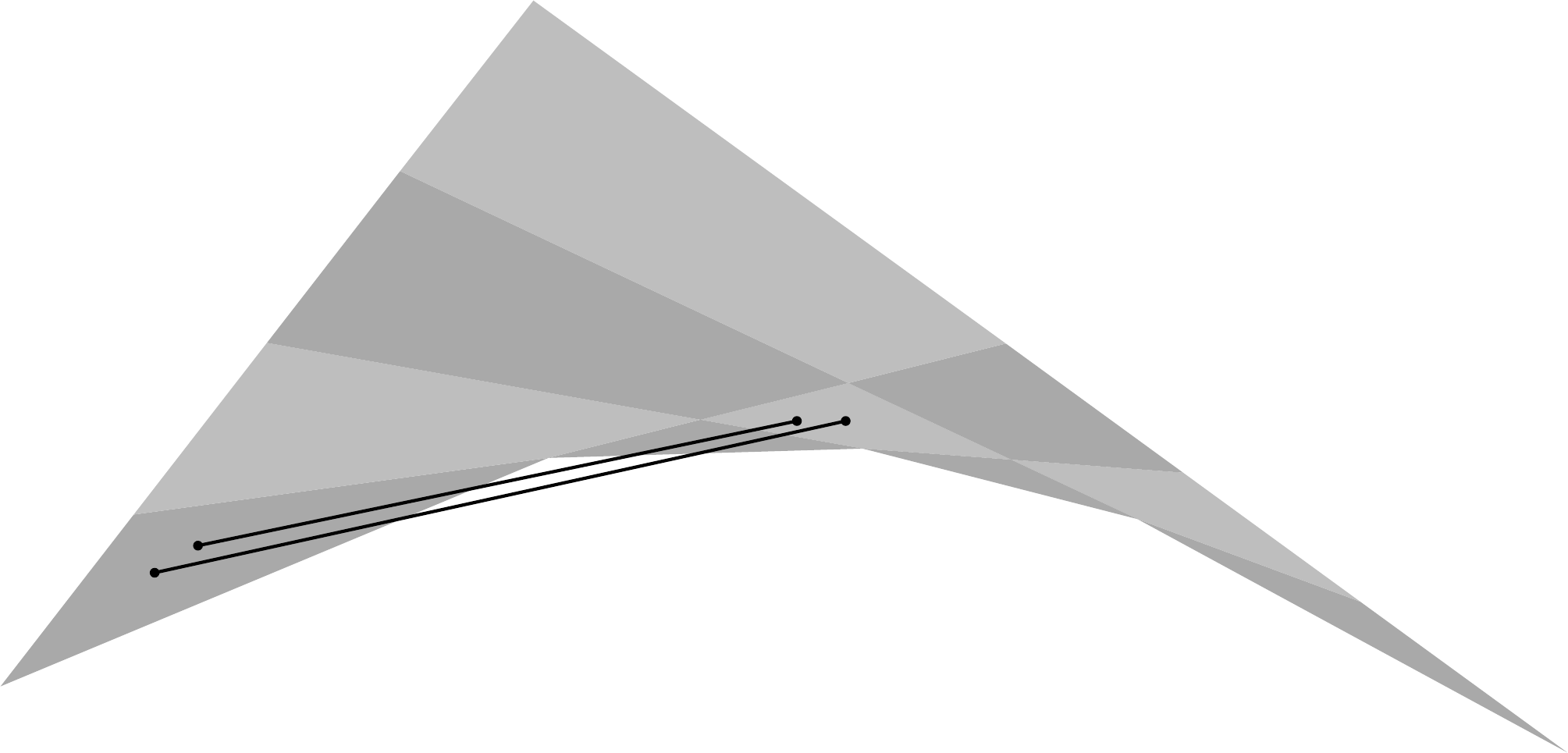}
	  \label{fig:coloring} \hspace{5mm}
	}
\subfigure[]{\includegraphics[scale=0.3]{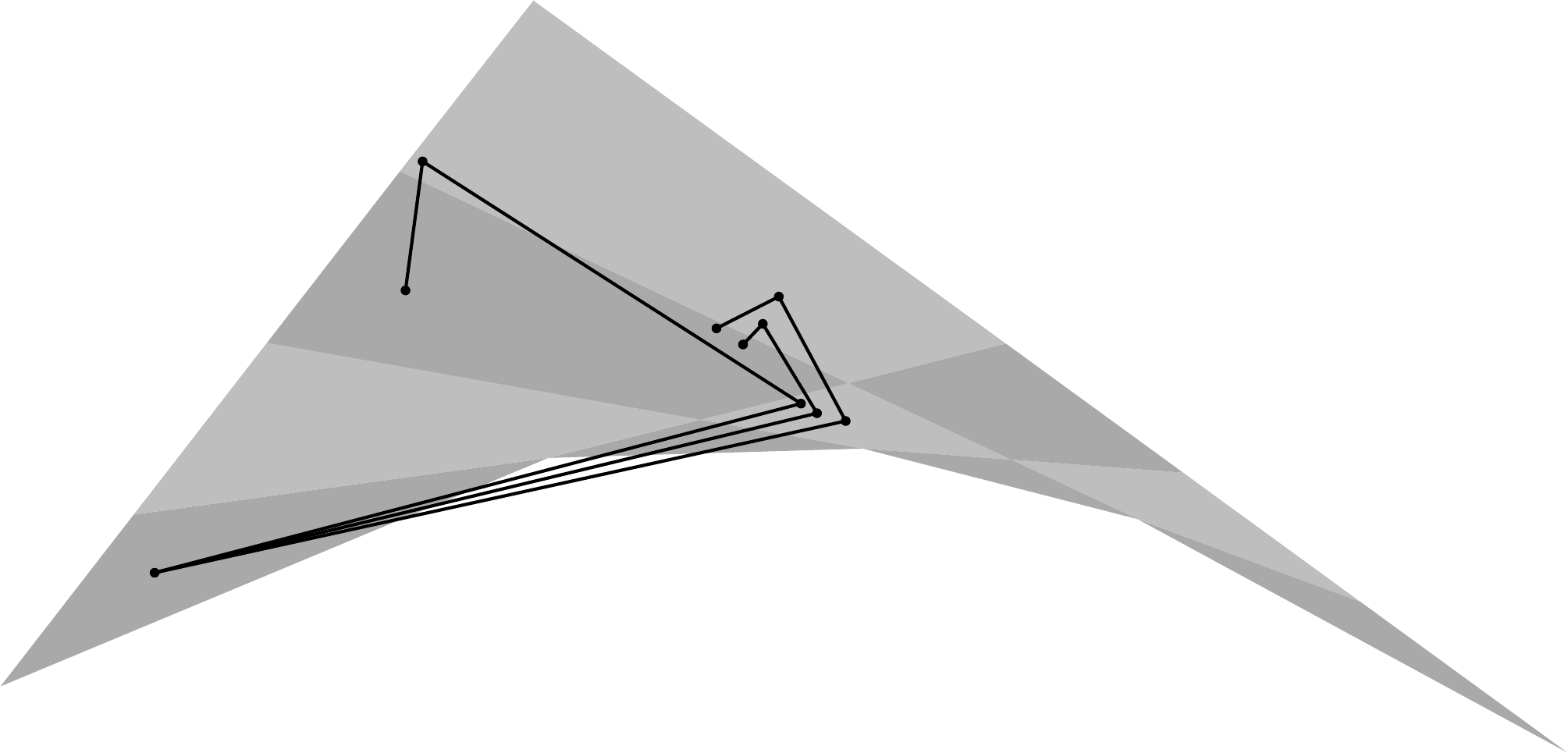}
	  \label{fig:uniform} \hspace{10mm}
	}
\subfigure[]{\includegraphics[scale=0.4]{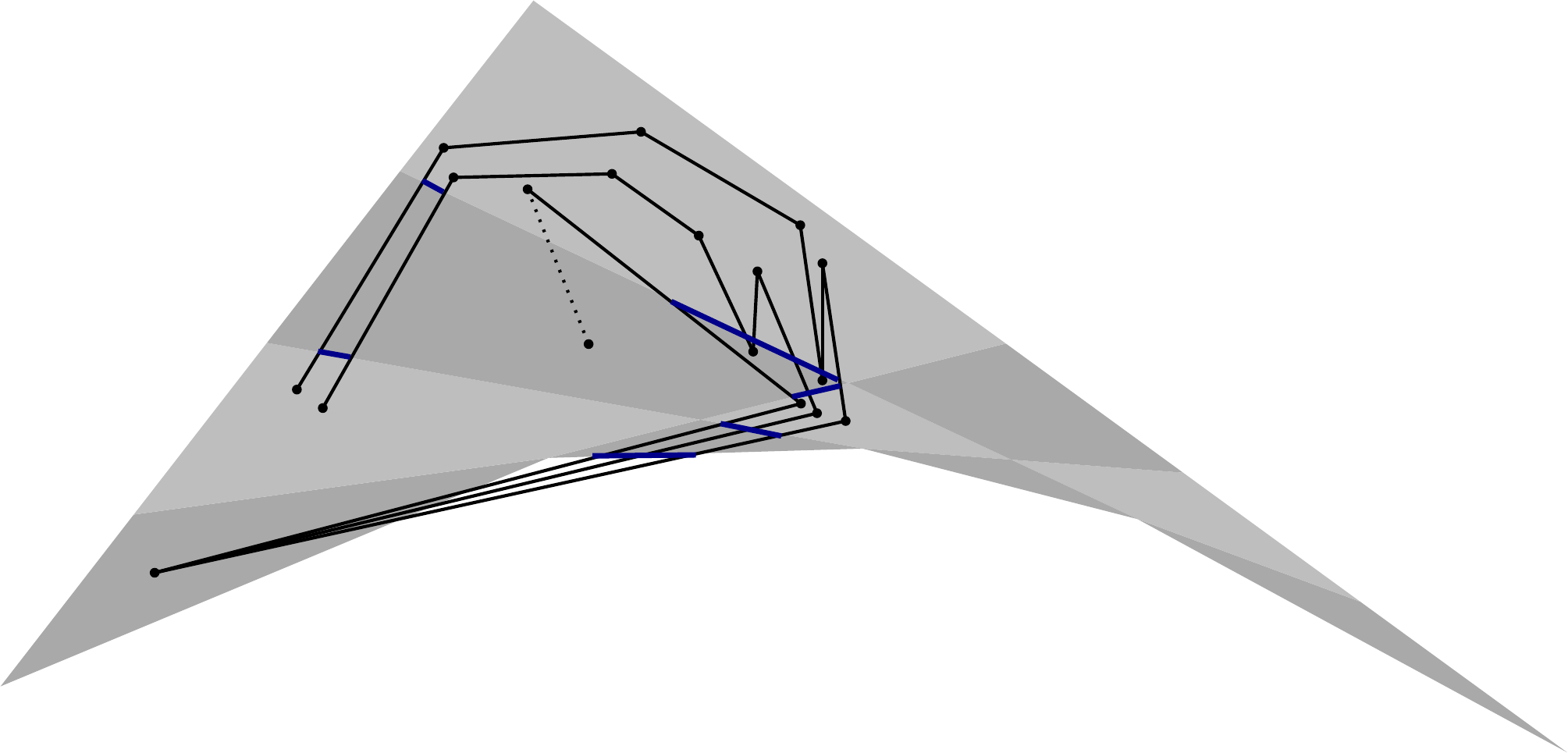}
	  \label{fig:superUniformity} \hspace{10mm}
	}
\caption{(a) Two edges having the same combinatorial type, (b) A set of paths having the same combinatorial type starting at the same vertex, (c) A set of two super uniform paths; the third path has the dotted edge violating the super uniformity; the doors are  indicated with the thick line segments.}
\end{figure}

By a standard perturbation argument we can assume that in any representation of $T$ as a geometric graph respecting $\iota$ none of the vertices of $T$ is represented by an intersection point
of two lines in $\mathcal{L}$ and no edge passes through such a point. Hence, every vertex belongs to exactly one region $R_{a,b}$.
Let $\overline{R_{a,b}}$ denote the closure of the convex hull of $R_{a,b}$. Note that $\overline{R_{a,b}}$ is always a polyhedron (possibly unbounded) with at most five sides
(see Fig.~\ref{fig:regions2} for an illustration).
We define the {\em combinatorial type} of an edge edge $e=\overrightarrow{uv}$  of $T$ to be a sequence of at most $c$ four-tuples $(a,b,x,y)$;  $a,b\in\{1,\ldots, c\};x,y\in\{0,1,2,3,4,5\}$ (see Fig.~\ref{fig:coloring} for an illustration), such that the $i$-th four-tuple stores in $a,b$ the $i$-th region $R_{a,b}$, which $e$ intersects on the way from $u$ to $v$ (the first region being the one containing $u$),
the values $x,y$ encode the sides of $\overline{R_{a,b}}$, through which $e$ enters and leaves $\overline{R_{a,b}}$ (0 for not entering or leaving the region).

Let $ct(e)$ denote the combinatorial type of $e\in E(T)$.
The {\em combinatorial type} of a path $v_1,\ldots, v_m$ is defined as the sequence of combinatorial types of its edges i.e. $(ct(\overrightarrow{v_1v_2}),\ldots, ct(\overrightarrow{v_{m-1}v_m}))$ (see Fig.~\ref{fig:uniform}). By $ct(P)$ we denote the combinatorial type of a path $P$.

We define the {\em color} of a vertex $v\in V$ as the natural number $c'$ such that $\iota(v)\in \mathcal{L}_{c'}$.
The {\em color type} of a path $v_1,\ldots, v_m$ is defined as an  $m$-tuple $(c_1,\ldots, c_m)$, such that $\iota(v_i)=\mathcal{L}_{c_i}$ for all $i=1,\ldots, m$. 
\begin{proposition}
\label{prop:regular}
For any $\Delta',d'>0$; $d'\leq d$, there exists $\Delta=\Delta(\Delta',d',c)$ such that $T=T(d,\Delta)$ contains a subtree $T_{d'}=T'(d',T)$ rooted at $r$ of depth $d$, such that
\begin{enumerate}[(i)]
\item
each non-leaf vertex $v\in V(T_{d'})$ has at least $\Delta'$ children of color $c'$ for each $c'=1,\ldots, c$;
\item for each vertex $v\in V(T_{d'})$ and
each color type $(c_1,\ldots, c_{m+1})$; $1\leq m\leq d'$, all paths in $T_{d'}$ starting at $v$ with the color type $(c_1,\ldots, c_{m+1})$  have the same combinatorial type.
\end{enumerate}
\end{proposition}

\begin{proof}
We prove the claim by induction on $d'$. For each $d',1\leq d'\leq d$ we inductively define an edge coloring $\chi_{d'}$  of $T$ leaving some edges uncolored, which encodes
for an edge $\overrightarrow{uv}$ the combinatorial types of all paths of length $d'$ having $\overrightarrow{uv}$ as the first edge.

For the base case $d'=1$. Let us color the edges of $T$ by their combinatorial types. Let $\chi_1$ denote this edge coloring.
We define $T_1$ as a subtree of $T$ in which each non-leaf vertex $u$ keeps its children $v$ of color $c'$, such that the color of the edge $uv$ occurs the most often among
the outgoing edges at $u$ joining $u$ with vertices of color $c'$. By the pigeon hole principle in $T_1$ each non-leaf vertex has still at least $\frac{\Delta}{cf(c,1)}$
children of color $c'$, for some function $f$ depending only on $c$ and $d'$. Hence, setting $\Delta=cf(c,1)\Delta'=\Delta(\Delta',1,c)$ finishes the base case.

For the inductive case, we assume that the claim holds up to $d'-1$ and we color  each edge $e=\overrightarrow{uv}$  of $T_{d'-1}$, so that $\chi_{d'-1}$ is defined for $\overrightarrow{vv'}$ where $v'$ is a child of $v$, by an ordered $c$-tuple of
 colors   $\chi_{d'}(\overrightarrow{uv})=(\chi_{d'-1}(\overrightarrow{vv_1}),\ldots, \chi_{d'-1}(\overrightarrow{vv_c}))$, where $v_{c'}\in T_{d'-1}$ and $\iota(v_{c'})\in \mathcal{L}_{c'}$.
 Our definition of $\chi_{d'}$ is not ambiguous, since  for all $c'\in \{1,\ldots ,c\}$ the color $\chi_{d'-1}(\overrightarrow{vv_{c'}})$ is the same for all children $v_{c'}$ of $v$ in $T_{d'-1}$, such that $\iota(v_{c'})\in \mathcal{L}_{c'}$.
 Note that the coloring $\chi_d'$ encodes the combinatorial types of paths of length $d'$ having $\overrightarrow{uv}$ as the first edge.

 Similarly, as in the base case we define $T_{d'}$ as a subtree of $T_{d'-1}$ in which each non-leaf vertex $u$ keeps its children $v$ of color $c'$, such that the color defined by $\chi_{d'}$ of the edge $\overrightarrow{uv}$ occurs the most often among the outgoing edges at $u$ joining $u$ with vertices of color $c'$. A vertex $u$ also keeps its children $v$ so that the edge $\overrightarrow{uv}$ is uncolored
 by $\chi_{d'}$.
By the pigeon hole principle in $T_{d'}$ each non-leaf vertex keeps at least $\frac{1}{f(c,d')}$ fraction of its children of color $c'$.
Moreover, if $T_{d'}$ contains a vertex $u$, and two paths $P_1=uv_1\ldots$ and $P_2=uv_2\ldots$ starting at $u$ having the same
color type, but not the same combinatorial type, then two edges $\overrightarrow{uv_1}$ and $\overrightarrow{uv_2}$ would be colored differently by $\chi_{d'}$ (by induction hypothesis).
 Thus, setting $\Delta=\Delta(f(c,d')\Delta',d'-1,c)=\Delta(\Delta',d',c)$ finishes the inductive case.
\end{proof}

We call a set of internally disjoint paths starting at the same vertex having the same length and combinatorial type {\em uniform}.
Let $\mathcal{P}$, $|\mathcal{P}|\geq 2$, denote a set of uniform paths in $T_{d'}=T'(d',T)$  of length $d''$ starting at $v\in V$.
We use the uniform set of paths $\mathcal{P}$ to define a special set of its subpaths which gives rise to a structure behaving
uniformly with respect to our regions $\overline{R_{a,b}}$.
First, let us introduce a couple of notions.

Let $\mathcal{P}'\subseteq\{P'\subseteq P| \ P\in \mathcal{P}, \ v\in P' \}$. Thus, $\mathcal{P}'$ is a set of subpaths of the paths in $\mathcal{P}$ with the same starting vertex.
We define the $i$-th {\em visited region} of $\mathcal{P}'$ to be the
$i$-th region $\overline{R_{a,b}}$ that we visit  (we count also revisits of the same region, see Figure~\ref{fig:visitingRegions}) when traversing a path in $\mathcal{P}'$ from $v$, the region containing $v$ being the $0$-th region. Note that a region $\overline{R_{a,b}}$ can be the $i$-th visited region for more than one $i$. The definition is correct by the fact that $\mathcal{P}$ is uniform.
We define the $i$-th {\em point of entry} of a path $P$ in $\mathcal{P}'$, as its $i$-th intersection point with the boundary of a region of the form $\overline{R_{a,b}}$,
in which we enter such a region, when  traversing the paths in $\mathcal{P}'$ from $v$.
Here, we ignored intersection points with the boundary of a region $\overline{R_{a,b}}$, in which we do not enter such region when  traversing the paths in $\mathcal{P}'$ from $v$.
 We define the $0$-th point of entry to be $v$.
We define the $i$-th {\em segment} of a path $P$ in $\mathcal{P}'$ to be the subpath of $P$ (in the topological sense) having the $i$-th and $(i+1)$-st point
of entry on $P$ as the endpoints. The $i$-th segment is not defined for the paths without the $(i+1)$-st point
of entry.
We define the $i$-th {\em door} of the set  $\mathcal{P}'$ to be the convex hull of the $i$-th points of entry of the paths in $\mathcal{P}'$.

We want the set of paths $\mathcal{P}$ to define an area that entraps some paths of $\mathcal{P}$ in its interior.
However, $\mathcal{P}$ as defined above is not 'uniform enough' for this purpose and we need to introduce a combinatorially 'more uniform' set of paths that gives rise to the required structure.
In the mentioned 'more uniform' set of paths the paths not only visit the same regions in the same order, but when leaving a region they always 'turn to the same side'
or return to the area that entraps the initial pieces of the paths in $\mathcal{P}$.

Ref. to Fig.~\ref{fig:superUniformity}.
A set of paths  $\mathcal{P}_{d''}$, $d''\ge0$,  is called \emph{super uniform} if it can be constructed from $\mathcal{P}$ by the following procedure.

The set $\mathcal{P}_0$ contains just one trivial path consisting of the single vertex $v$.

Having defined $\mathcal{P}_{j}$ we define $\mathcal{P}_{j+1}$. Let $\mathcal{P}_{j}^l$ denote the subset of $\mathcal{P}_j$ containing its longest paths. Let $\overline{R_{a_j,b_j}}$  denote the region containing the last points of the paths in $\mathcal{P}_j^l$.
Let $\mathcal{P}_{j+1}'=\{Pv'| \ P\in \mathcal{P}_j^l, \ Pv'\subseteq P'\in \mathcal{P}\}$.
Let $i$ be the maximal integer such that path(s) in $\mathcal{P}_j$ define at least one $i$-th point of entry. Let $l$ be the line containing the $i$-th door of $\mathcal{P}$.
If the last edges of paths in $\mathcal{P}_{j+1}'$ do not intersect the side of  $\overline{R_{a_j,b_j}}$ containing the $i$-th door of $\mathcal{P}$ we set $\mathcal{P}_{j+1}=\mathcal{P}_{j+1}'\cup(\mathcal{P}_j\setminus \mathcal{P}_j^l)$.
Otherwise, we let $l'$ denote a connected component of $l\setminus (i{\rm-th \  door})$ and  define $\mathcal{P}_{j+1}^c$ and $\mathcal{P}_{j+1}^d$, respectively, to be the subset of $\mathcal{P}_{j+1}'$ containing the paths
whose last edges intersect $l'$ and the $i$-th door, respectively.
We set either
\begin{enumerate}[(i)]
\item
$\mathcal{P}_{j+1}=\mathcal{P}_{j+1}^c\cup(\mathcal{P}_j\setminus \mathcal{P}_j^l)$ or
\item
 $\mathcal{P}_{j+1}=\mathcal{P}_{j+1}^d\cup(\mathcal{P}_j\setminus \mathcal{P}_j^l)$ or
 \item
 $\mathcal{P}_{j+1}=(\mathcal{P}_{j+1}'\setminus(\mathcal{P}_{j+1}^c \cup \mathcal{P}_{j+1}^d))\cup(\mathcal{P}_j\setminus \mathcal{P}_j^l)$.
\end{enumerate}
We have the following.

\begin{proposition}
\label{prop:maximum}
From a uniform set $\mathcal{P}$ of paths of length $d'$ we can construct a super uniform set with at least
${3^{-d'}}|\mathcal{P}|$ paths of length $d'$.
\end{proposition}
\begin{proof}
The claim follows if we adjust the definition of the super uniform set of paths as follows.
If $|\mathcal{P}_{j+1}^c|\geq \frac{1}{3}|\mathcal{P}_{j+1}'|$ we set $\mathcal{P}_{j+1}=\mathcal{P}_{j+1}^c\cup(\mathcal{P}_j\setminus \mathcal{P}_j^l)$.
If $|\mathcal{P}_{j+1}^c|< \frac{1}{3}|\mathcal{P}_{j+1}'|$  and $|\mathcal{P}_{j+1}^d|\geq \frac{1}{3}|\mathcal{P}_{j+1}'|$ we set $\mathcal{P}_{j+1}=\mathcal{P}_{j+1}^d\cup(\mathcal{P}_j\setminus \mathcal{P}_j^l)$.
Otherwise, we set $\mathcal{P}_{j+1}=(\mathcal{P}_{j+1}'\setminus(\mathcal{P}_{j+1}^c \cup \mathcal{P}_{j+1}^d))\cup(\mathcal{P}_j\setminus \mathcal{P}_j^l)$.
\end{proof}


\bigskip

\begin{figure}[htp]
  \centering
  \subfigure[]{\includegraphics[scale=0.5]{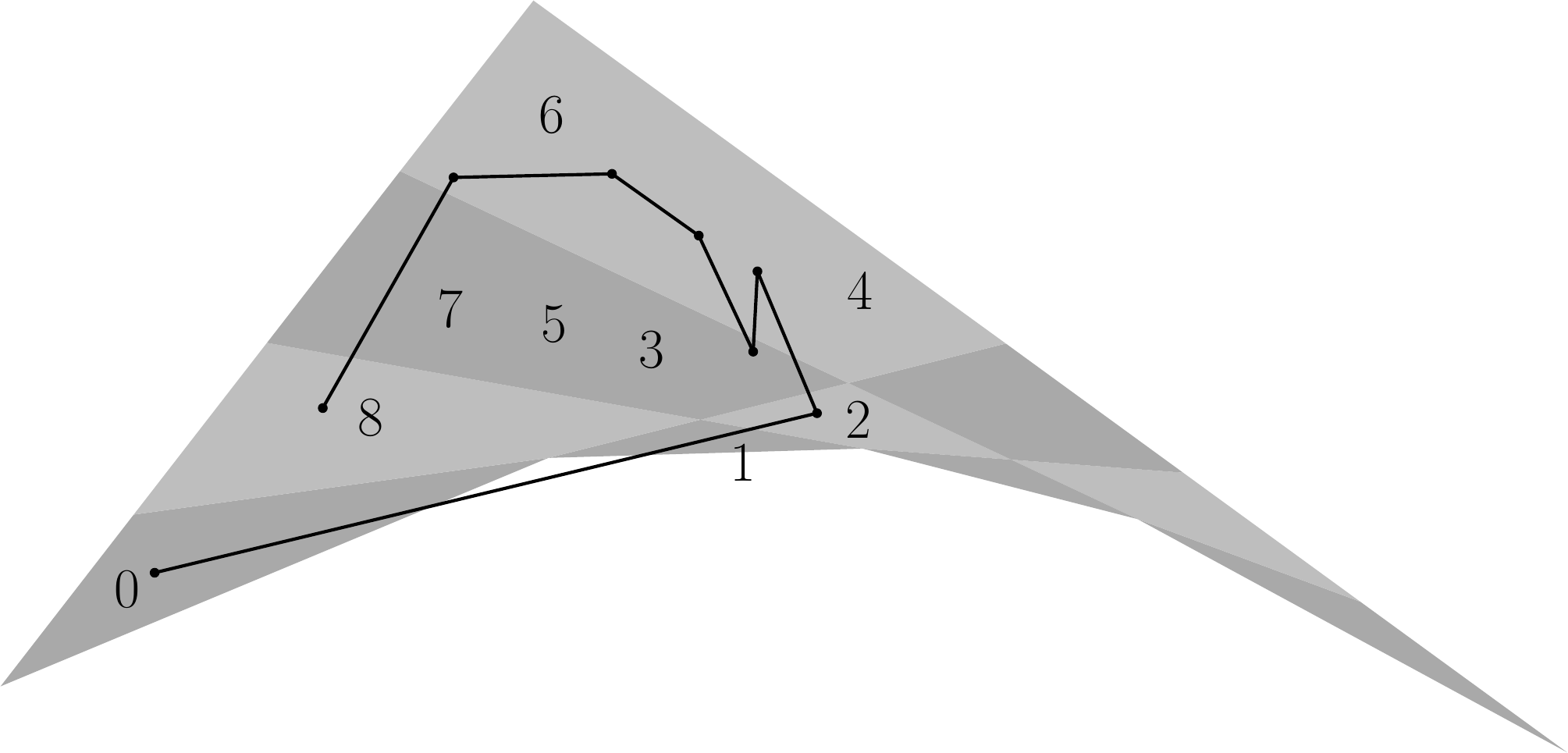}
	  \label{fig:visitingRegions} \hspace{10mm}	}
   \subfigure[]{\includegraphics[scale=0.5]{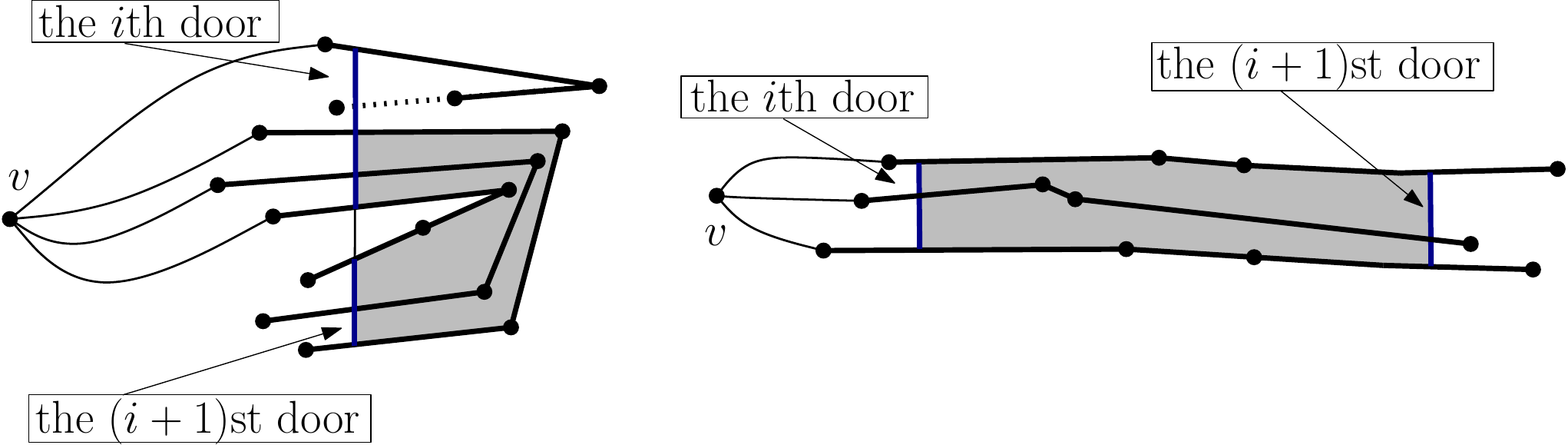}
	  \label{fig:delimiting}
	}\hspace{1cm}
  \subfigure[]{\includegraphics[scale=0.5]{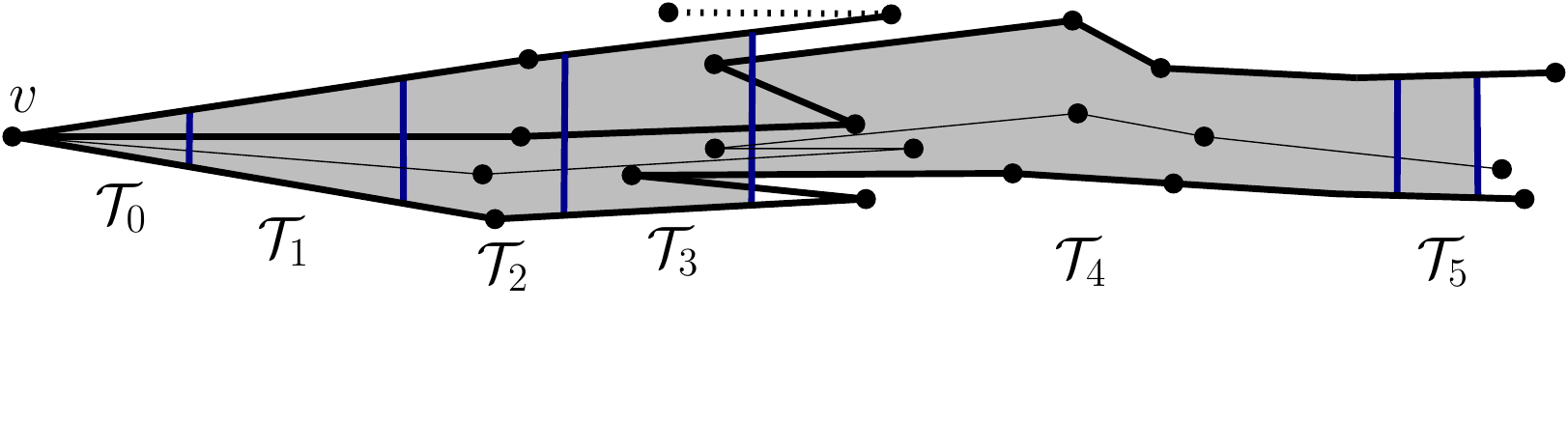}
	  \label{fig:superUniformity2} \hspace{10mm}
	}
\caption{(a) The visited regions; (b)  Tubus polygons of $\mathcal{P}$ are colored by grey; on the left the $i$th and $(i+1)$st doors belong to the
 same line; (c) Tubus of length 6; its tubus polygons (colored by grey) separated by vertical segment;
the thin path is central)}
\end{figure}

From now on we suppose that $\mathcal{P}$ is a set of super uniform paths starting at $v\in T_{d'}$ having vertices in $\mathcal{T}_{d'}$. 
Suppose that $i$-th segments of paths in $\mathcal{P}$ intersect $i$-th door exactly once.
Let $S_1,\ldots, S_{i'}$ denote the $i$-th segments of paths in $\mathcal{P}$ listed according to their appearance on the $i$-th (resp. $(i+1)-st$) door.
Let $s_1,\ldots, s_{i'}$ denote the intersection points of $S_1,\ldots, S_{i'}$, resp., with the $i$-th door.
A {\em tubus polygon} of $\mathcal{P}$ (see Fig.~\ref{fig:delimiting}) is the area
  bounded by $S_1, S_{i'}$, the line segment $s_1s_{i'}$, and the $(i+1)$-st door.
By a {\em tubus} $\mathcal{T}=\mathcal{T}(\mathcal{P})$ of $T_{d'}$
{\em starting} at $v$
 we understand the union of tubus polygons defined by $\mathcal{P}$. 
We define a {\em maximal tubus polygon} of a tubus as a tubus polygon not properly contained
in other tubus polygon of the tubus.
 Note that if two tubus polygons of the same tubus
are not internally disjoint then necessarily one of them contains the other.
We define the {\em length} of a tubus as the number of maximal tubus polygons it consists of.
Let $\mathcal{T}_i$ denote the  $i$-th, $0\leq i$, maximal tubus polygon (the 0th one is the one containing $v$) of $\mathcal{T}$
with respect to the order, in which the tubus polygons are visited for the first time as we traverse the paths in $\mathcal{P}$ from $v$.
We say that a longest path in $\mathcal{P}$ is {\em central} (see Fig.~\ref{fig:superUniformity2}) if none of its segments bounds a maximal
tubus polygon of $\mathcal{T}(\mathcal{P})$.

By Jordan Curve Theorem we have the following simple property of  $\mathcal{P}$.

\begin{proposition}
\label{prop:ordering}
Let $\mathcal{P}'\subseteq \mathcal{P}$ denote a set of internally disjoint paths of $\mathcal{P}$ starting at $v$.
Let $i$ be such that the $i$-th door of $\mathcal{P}$ are disjoint from the first $i-1$ segments of the paths in $\mathcal{P}$.
The ordering of the paths in $\mathcal{P}'$ according to the appearance of their $i$-th points of entry is the same (up to  reverse) as the ordering of the paths in $\mathcal{P}'$ according to the appearance of their first points of entry on the first door.
\end{proposition}

\begin{figure}[htp]
\centering
\includegraphics[scale=0.7]{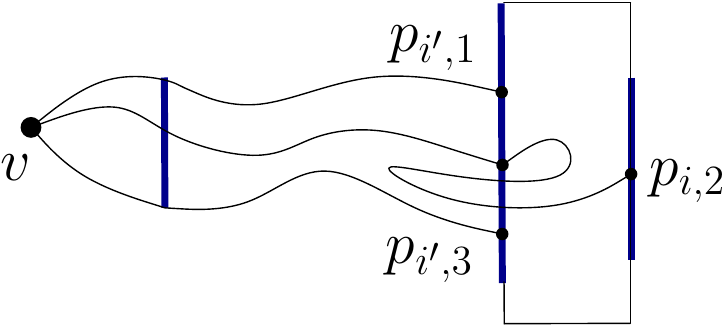}
\label{fig:prop33}
\caption{An illustration for the proof of Proposition~\ref{prop:ordering}.}
\end{figure}
\begin{proof}
The claim is trivial if $|\mathcal{P}|\leq 2$. Otherwise, let $P_1,P_2,P_3\in \mathcal{P}$, indexed according to  the appearance of their first  points of entry
on the first door. We prove the claim by induction on $i$. The base case, when $i=1$, is easy.

Let $i', 0<i'<i,$ denote the minimal $i'$ so that the $i'$-th segments of the paths in $\mathcal{P}$ intersect the $(i-1)$-st door.

Let $p_{j,1},p_{j,2}$ and $p_{j,3}$ denote the $j$-th point of entry of $P_1,P_2$ and $P_3$, respectively.
Let $\mathcal{C}$ denote the Jordan curve, which is the union of the part of the $i'$-th door between $p_{i',1}$ and $p_{i',3}$ and the subpaths of $P_1$ and $P_3$ starting at $v$ and ending at their respective $i'$-th points of entry.
By induction hypothesis, $\mathcal{C}$ bounds a region that contains the subpath of $P_2$ having $v$ and $p_{i',2}$ as the endpoints.
It follows that $p_{i-1,1},p_{i-1,2}$ and $p_{i-1,3}$ appear in this order on the $(i-1)$-st door, which in turn implies that
 $p_{i,1},p_{i,2}$ and $p_{i,3}$ appear in this order on the $i$-th door (see Fig.~\ref{fig:prop33}).
\end{proof}

Thus, by Proposition~\ref{prop:ordering} two longest paths (if they exist) of $\mathcal{P}$ up to the furthest door together with the last door of $\mathcal{T}$ form the boundary of a simply connected compact region containing all the maximal paths in $\mathcal{P}$ up to the last door.

The next lemma states an important property of tubuses, which is intuitively quite expectable and it says that
by letting $\Delta$ and $d'$ grow, the length of a longest tubus of $T_{d'}$ grows as well.

\begin{lemma}
\label{lemma:tubusPolygon}
Let $c\geq 20$.
For every $k>0$  there exists  $d'=d'(k)$ and $\Delta=\Delta_0(k,c,p)$  such that for every $d\geq d'$, and every vertex $v\in V(T_{d'})$ at distance at most $d-d'$ from the root of  $T_{d'}=T'(d',T(d,\Delta))$ and $C\in {\{1,\ldots, c\} \choose 20}$ there exists a tubus $\mathcal{T}$ of  $T_{d'}$  of length at least $k$ starting at $v$ such that $\mathcal{T}$ has the vertices ($\not= v$)
of its defining  paths  colored by elements of $C$ and has at least $p$ internally disjoint (super uniform) paths of length $d'$.
\end{lemma}

\begin{proof}
Observe that it is enough to prove the lemma for $c=20$, as for higher $c$ we can take $\Delta=\Delta_0(k,20,p)\frac{c}{20}$.
Then for every non-leaf vertex $u$ and $c'\in \{1,\ldots, c\}$, at least $\frac{\Delta_0(k,20,p)}{20}$  of children of $u$ are mapped to the class $\mathcal{L}_{c'}$. Hence, the lemma follows by considering the maximal subtree $T'$ of $T$ rooted at $r$ whose vertices ($\not=r$) are colored by elements of $C$.

First, we assume that $d=d'$.
For the sake of contradiction let $k_0$ denote the maximum length of  a tubus of  $T_{d'}$ with $p3^{5k_0+7}3^{5k_0+7+d_0}+2$ internally pairwise disjoint paths of length $d'$
in its defining set  for sufficiently large $d'=d_0'$ and $\Delta$. By taking a sufficiently big $\Delta$, $k_0>0$.
In what follows we show that for $d'=d_0'+5k_0+7$ the tree $T_{d'}$ has to contain  a tubus of length $k_0+1$ having at least $p$ internally disjoint paths of length $d'$ in its
defining set of super uniform paths, which is a desired contradiction. Throughout the proof of the lemma all the edges and vertices are meant to be in $T_{d'}$.


Let $\Delta=\Delta(\Delta, d', 20)$, where $d'=d_0'+5k_0+7$ and $\Delta(\Delta, d', 20)$ is as in Proposition~\ref{prop:regular}.
Let $\mathcal{T}(\mathcal{P})$ denote a maximum length tubus of $T_{d'}=T'(d',T(d,\Delta))$ with $p3^{5k_0+7}3^{5k_0+7+d_0}+2$ internally disjoint maximal paths in the defining set so that the paths in $\mathcal{P}$ start at $r$.  We can assume that $\mathcal{T}$ has the length of $k_0$, as otherwise we are done.
 Let $V_i'$ denote the set of the $i$-th vertices
of the paths in $\mathcal{P}$.
Let $\mathcal{T}_{k_0}$ denote the region $\overline{R_{a,b}}$ containing $V_{i}'$ which is not contained in a tubus polygon.
This is the region through which $\mathcal{T}$ could
be possibly prolonged.
Let $c$ and $c-1$ (resp. $c$)
 denote the color classes  (resp. class) of the lines that intersect $\mathcal{T}_{k_0}$.

Let $V_0\subseteq V_{d_0'}'$ consists of the vertices on the central paths of $\mathcal{P}$.
Let $C_0\subseteq \{1,\ldots, c-2\}$ denote the set of size at least $(c-4)/2$ such that the edges between the vertices in $V_0$ and their children  having the colors in $C_0$,
intersect the same side of the region $\overline{R_{a,b}}$,  which the vertices of $V_{d_0'}'$ belong to.
Since we are not allowed to prolong the tubus
$\mathcal{T}$, if $V_{d_0'}'$ is contained in $\mathcal{T}_{k_0}$, all the edges connecting $V_{d_0'}'$ with their children
 have to cross the last door. Otherwise, they have to cross one of the two doors, or stay inside the current maximal tubus polygon, which intersects the lines in at most two classes $\mathcal{L}_{c'}$.

In what follows we define triples $(V_x,C_x,C_x')$, s.t. $V_x\subseteq V(T_{d'})$, $C_x'\subseteq C_x \subseteq \{1,\ldots, c-2\}$,  for $x=1,\ldots,5k_0+7$, giving rise to the subtrees contained (by our assumption) in $\mathcal{T}$.

Let $V_1$ denote the set of children of the vertices in $V_{0}$ having the same color $c'\in C_0$.
We denote by $C_1\subseteq \{1,\ldots, c-2\}$ a set of size at least $(c-6)/2$ such that the edges between the vertices $V_{1}$ and their children
 having the colors in $C_1$ intersect the same side of a region $\overline{R_{a,b}}$ that $V_{1}$ belongs to, and also we require that $C_1$ does not
 contain any $c'$ such that the lines in $\mathcal{L}_{c'}$ intersect the regions of the form $\overline{R_{a,b}}$ containing $V_0$.
Let us pick the maximal subset $C_1' \subseteq C_1$ so that the edges from $V_1$ to their children having the colors in $C_1'$, intersect the most number of regions of the form $\overline{R_{a,b}}$. Note that $|C_1'|\leq 2$.

\begin{figure}[htp]
\centering
\includegraphics[scale=1.2]{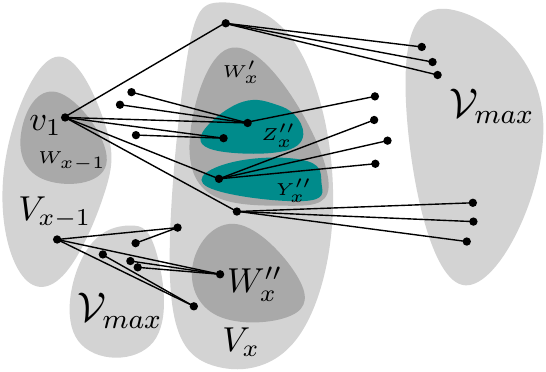}
\label{fig:lemma34}
\caption{Constructing $V_{x+1}$. The figure is slightly misleading since $V_{max}$ belongs, in fact, to a single tubus polygon.}
\end{figure}

In general, having defined the triple  $(V_x,C_x,C_x')$ we define the triple $(V_{x+1}, C_{x+1}, C_{x+1}')$ as follows.

Ref. to Fig~\ref{fig:lemma34}.
For $v_1\in V_{x-1}$ we define $\mathcal{T}^{x+1}(v_1)$ to be the tubus $\mathcal{T}(\mathcal{P}_{v_1})$, where $\mathcal{P}_{v_1}$ is the set of
super uniform paths obtained by applying Proposition~\ref{prop:maximum} to the set of all paths of the form $v_1v_2v_3$, such that $v_2\in V_{x}$ and
$v_3\in V_{max}$, where $V_{max}$ is the set of children  of the vertices in $V_{x}$ having the same color $c'\in C_{x}'$ (chosen arbitrarily).
Let $W_{x-1}$ be the subset of $V_{x-1}$ consisting of all $v_1$ for which the doors defined by $\mathcal{T}^{x+1}(v_1)$ are pairwise disjoint.
Let $W_{x}'$ denote the subset of $V_{x}$ consisting of all vertices in $V_x$ on the central paths of $\mathcal{T}^{x+1}(v_1)$ for all $v_1\in W_{x-1}$.
Let us arbitrarily choose $c'\in C_x\setminus C_x'$. We denote by $V_{v_2}$ the set of children of the vertex $v_2\in W_x'$ having the color $c'$.
We denote by $V_{v_2}'$ the subset of $V_{v_2}$ so that $v_2v_2'$, $v_2'\in V_{v_2}'$, does not intersect the door of $\mathcal{T}(v_1)$ intersected by the line segment $v_1v_2$, where $v_1$ is the parent of $v_2$. We say that $v_2\in W_{x}'$ is of type (i) if $|V_{v_2}'|\ge |V_{v_2}\setminus V_{v_2}'|$ and of type (ii) otherwise.
Let $Y_{x}''$ denote the subset of $W_x'$ of the vertices of type (i), and
let $Z_{x}''$ denote the subset of $W_x'$ of the vertices of type (ii).
Let $W_{x}''$ denote the subset of $V_{x}$ consisting of the second vertices on the central paths of $\mathcal{T}^{x+1}(v_1)$ for all $v_1\in V_{x-1} \setminus W_{x-1}$.
Let $A_{x},B_{x}$ and $C_{x}$ denote the set of ancestors of $Y_{x}'',Z_{x}''$ and $W_{x}''$, resp., in $V_0$.

If $|A_{x}|\ge \max\{|B_{x}|, |C_{x}|\}$, let $V_{x+1}=\bigcup_{v_2\in Y_{x}''}V_{v_2}'$.
Otherwise, if $|C_{x}|\ge|B_{x}|$,  $V_{x+1}=V_{max}$, and if $|C_{x}|<|B_{x}|$, $V_{x+1}=\bigcup_{v_2\in Z_{x}''}(V_{v_2}\setminus V_{v_2}')$.
Thus, the second and the third case, i.e. $|A_{x}|< \max\{|B_{x}|, |C_{x}|\}$, correspond to the situation when $V_{x+1}$ ends up in a tubus polygon of $\mathcal{T}$ situated between tubus polygons $\mathcal{T}_{x}$ and $\mathcal{T}_{x+1}$.

We denote $C_{x+1}\subseteq \{1,\ldots, c-2\}$ a subset of size at least $(c-14)/2$ such that the edges between the vertices $V_{x+1}$ and their children
having the color in $C_{x+1}$ intersect the same side of a region $\overline{R_{a,b}}$ that $V_{x+1}$ belongs to, and we also require that $C_{x+1}$ does not
 contain any $c'$ such that the lines in $\mathcal{L}_{c'}$ intersect the region of the form $\overline{R_{a,b}}$ containing $V_{x}, V_{x-1}, V_{x-2},V_{x-3}, V_{x-4}$ or $V_{x-5}$  (whenever they are defined for our $x$, of course).
Let us pick $C_{x+1}' \subseteq C_{x+1}$ so that the edges from $V_{x+1}$ to their children having the colors in $C_{x+1}'$, intersect the most number of regions of the form $\overline{R_{a,b}}$.

Since $V_0\ge p3^{5k_0+7}3^{5k_0+7+d_0}$, by the construction of $V_i$-s, there exists a set of size of at least $p3^{5k_0+7+d_0}$
of internally pairwise disjoint uniform  paths of the form $r=v_0v_1,\ldots, v_{d_0'},\ldots ,v_{d'}$, $v_i\in V_i'$, for $0<i\leq d_0'$, $v_i\in V_{i-d_0}$, for $i>d_0'$.
By Proposition~\ref{prop:maximum} a set of uniform  paths of the form $r=v_0v_1,\ldots, v_{d_0'},\ldots ,v_{d'}$, $v_i\in V_i'$, for $0<i\leq d_0'$, $v_i\in V_{i-d_0}$, for $i>d_0'$, gives
rise to a tubus with at least $p$ internally disjoint paths (of length $d'$). In what follows we show that the length of this tubus has to be at least $k_0+1$, which is a contradiction.

%

Let  $V_{x} \subseteq \mathcal{T}_{i'}$, $V_{x+1} \subseteq \mathcal{T}_i$, $V_{x+2} \subseteq \mathcal{T}_j$, $x\geq 0$.
 We let $I_x=(l_x,u_x)$ denote an interval, such that $I_0=(0, k_0)$.
We claim the following:

\begin{enumerate}[(i)]
\item
If $i'<i<j$,  vertices of $\cup_{y=x}^{5k_0+7}V_y$ have to belong to the union of tubus polygons $\cup_{y=i'}^{u_{x}}\mathcal{T}_{y}$.
We set  $I_{x+1}=(i',u_x)$.
\item
If $i'<i$ and $j<i$,  vertices of $\cup_{y=x}^{5k_0+7}V_y$ have to belong either a) to the union of tubus polygons $\cup_{y=l_{x}}^{i}\mathcal{T}_{y}$, if
$|A_{x}|\geq \max\{|B_x|,|C_x|\}$, or b) to the union of polygons $\cup_{y=i'}^{u_{x}}\mathcal{T}_{y}$, otherwise.
We set a) $I_{x+1}=(l_x,i)$ or b) $I_{x+1}=(i',u_x)$, accordingly.
\item
If $i'>i>j$,  vertices of $\cup_{y=x}^{5k_0+7}V_y$ have to belong to the union of tubus polygons $\cup_{y=l_{x}}^{i'}\mathcal{T}_{y}$.
We set  $I_{x+1}=(l_x,i')$.
\item
If $i'>i$ and $i<j$,  vertices of $\cup_{y=x}^{5k_0+7}V_y$ have to belong either a) to the union of tubus polygons  $\cup_{y=i}^{u_{x}}\mathcal{T}_{y}$, if
$|A_{x}|\geq \max\{|B_x|,|C_x|\}$, or b) to the union of polygons $\cup_{y=l_{x}}^{i'}\mathcal{T}_{y}$, otherwise.
We set a) $I_{x+1}=(i,u_x)$ or b) $I_{x+1}=(l_x,i')$, accordingly.
\end{enumerate}
\begin{figure}
  \centering
  \subfigure[]{
  \includegraphics[scale=0.7]{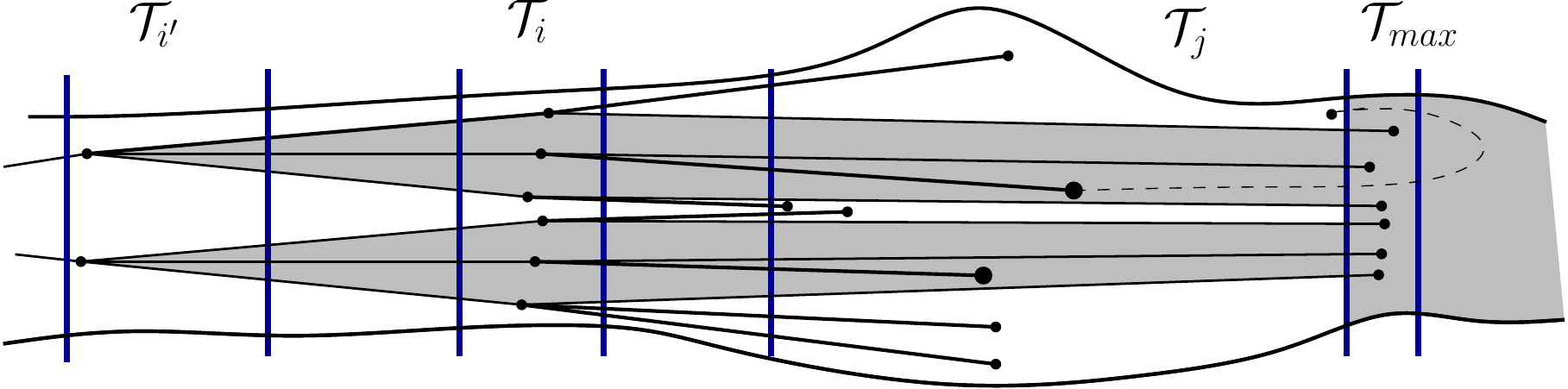}
  \label{fig:firstLemma}}
   \subfigure[]{
  \includegraphics[scale=0.6]{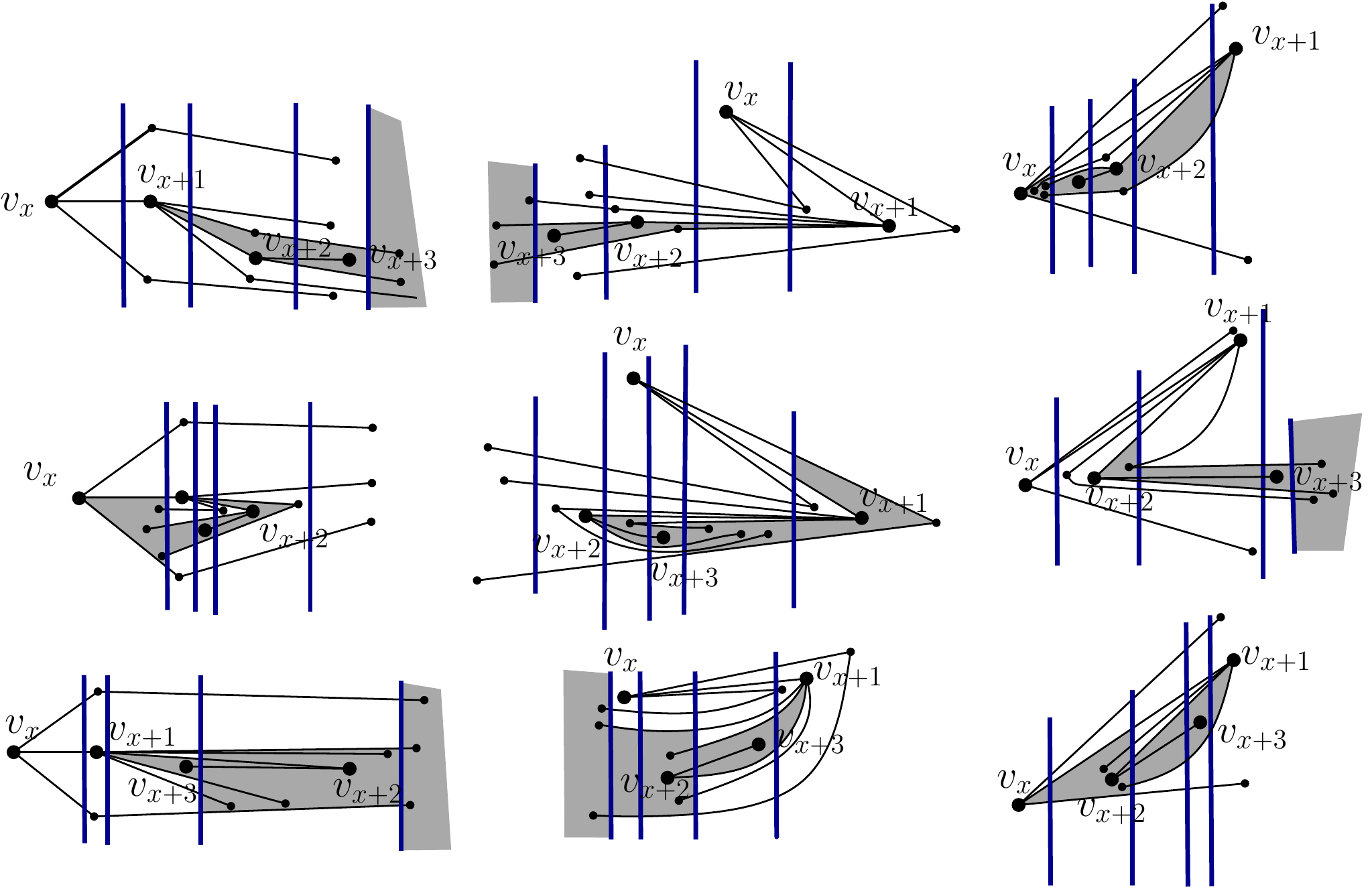}
  \label{fig:firstLemma2}}

  \caption{(a) Illustration for  case (i); $\mathcal{T}_{max}$ is the tubus polygon intersecting the classes of lines in $\mathcal{L}_{c'}$, for
  $c'\in C_{x+1}'$, the descendants of the bold vertices have to stay in the respective gray regions. (b) Two consecutive steps in the inductive case, the descendants of the bold vertices have to stay in the respective darkgray regions.}
\end{figure}

 Suppose for a while that the conditions (i)-(iv) hold.

Then it must be that the interval $I_{x+1}$ is always contained in $I_{x}$.
Moreover, we show that the intervals $I_{x+1}$ and $I_{x+5}$ are never equal, which means that $I_x$ is shorten by at least one during any five consecutive inductive steps:
 By the construction $V_x,V_{x+1}, V_{x+2}$  and $V_{x+3}$, respectively,  belong to four different tubus polygons $\mathcal{T}_{x_0}, \ldots, \mathcal{T}_{x_3}$, respectively,
 of $\mathcal{T}$.
 By a simple case analysis it follows that if $I_{x+1}$ is the same as $I_{x+3}$, both $l_{x+3}$ and $u_{x+3}$ belong to $\{x_0,\ldots x_3\}$.
 Hence, $I_{x+5}$ has to be different from $I_{x+3}$.
Thus, we have no region to accomodate $V_{5k_0+7}$ (contradiction). It is left to prove (i)-(iv).
 \\

%
%

The proof which follows is rather straightforward by using Proposition~\ref{prop:ordering} 
 and a simple fact that  an edge cannot cross a line twice.
 Note that the cases (i) and (iii) (resp. (ii) and (iv)) are symmetric. Thus, the omitted part of the proof below can be easily filled in by the reader.
  Ref to Figs.~\ref{fig:firstLemma} and~\ref{fig:firstLemma2}:

 In Fig.~\ref{fig:firstLemma2} the first, second and third column depicts two consecutive inductive steps: the $x$-th, $(x+1)$-st step,
 when in the $x$-th step case (i), case (ii) a), and case (ii) b), resp., applies.
 Then the first, second and third row depicts two consecutive inductive steps: the $x$-th, $(x+1)$-st step,
 when in the $(x+1)$-st step case (i) or (iii), case (ii) a) or (iv) a), and case (ii) a) or (iv) b), resp., applies.
 Then the vertices $v_x,\ldots, v_{x+3}$, respectively, belong to $V_x,\ldots, V_{x+3}$, respectvively.

 We prove the claims (i)-(iv) by induction on $x$.
 Observe that it is enough to prove that $V_x,V_{x+1},V_{x+2}\subseteq\cup_{y\in I_{x+1}}\mathcal{T}_y$, since then
 it follows that $I_{x+1}\subseteq I_x$ for $x\geq 0$.
  The base case is easy to check.

 For the inductive case, by induction hypothesis the vertices in $V_{x},V_{x+1}, V_{x+2}$, for some fixed $x\geq 0$, are contained in $\cup_{y\in I_{x+1}}\mathcal{T}_y$.
 Hence, $I_{x+2}$ is contained in $I_{x+1}$, and by an easy inspection of all nine cases from Fig.~\ref{fig:firstLemma2} we also have that $V_{x+1}, V_{x+2}$ are contained in $\cup_{y\in I_{x+2}}\mathcal{T}_y
 $.
 Thus, it is enough to check that $V_{x+3}$ is always contained in $\cup_{y=l_{x+2}}^{u_{x+2}}\mathcal{T}_{y}$.

 By an inspection of nine possibilities from Fig.~\ref{fig:firstLemma2} of how two consecutive steps in the induction might look like, we easily rule out
 the possibility that some of the vertices $V_{x+3}$ for $x>0$ belong to $\cup_{y=0}^{l_{x+2}-1}\mathcal{T}_{y}$
 (resp. $\cup_{y=u_{x+2}+1}^{k_0}\mathcal{T}_{y}$).
 In fact, for all nine considered possibilities except the following three:  case (i) followed by case (i), case (ii) a) followed by case (iii), and
  case (ii) b) followed by case (iv) a), just by considering the two consecutive steps, $V_{x+3}$ has to belong to
  $\cup_{y=l_{x+2}}^{u_{x+2}}\mathcal{T}_{y}$.
Thus, in case (i) followed by case (i), and in case (ii) b) followed by case (iv) a),
 we still have to prove that the vertices $V_{x+3}$ for $x>0$ cannot belong to $\cup_{y=u_{x+2}+1}^{k_0}\mathcal{T}_{y}$.
 Similarly, it is also left to prove that in case (ii) a) followed by case (iii) the vertices $V_{x+3}$ for $x>0$ cannot belong to $\cup_{y=0}^{l_{x+2}-1} \mathcal{T}_{y}$.

 In each of the above three situations we can proceed by distinguishing the case, that set the value $u_x$ the most recently before the $x$-the step.
 Note that $u_x$ could be set either in case (ii) a), (iii), or (iv) b).
 In fact, the argument is almost the same for every situation, by observing that instead of case (ii) a) followed by case (iii), we can by symmetry consider
 case (iv) a) followed by case (i).

 First, suppose that $u_x$ was the most recently set in $I_{x'+1}$ such that in the $x'$-th step case (iii) applies. Let $y'$ be such that $\mathcal{T}_{y'}$ contains the children of $V_{x'+1}$ of color in $C_{x'+1}'$. Observe that the $x'$-th step has to be followed by
 a step in which case (iv) a) applies. Indeed, otherwise $u_x$ would be changed. This step is then followed by a sequence of steps looking (in general) as follows:
  case (i), case (i),$\ldots$, case (i), case (ii) b), case (iv) a), case (i), case (i),$\ldots$, case (i), case (ii) b), case (iv) a),$\ldots$, with
  the possibility of the omission of any of cases (i).
  Now, it is easy to see that $V_{x+3}$  cannot belong to $\cup_{y=u_{x+2}+1}^{k_0} \mathcal{T}_{y}$: For all $V_{x''}\subset \mathcal{T}_{y''}$; $x+3>x''>x'+1$ we have
  $y''>y$, where $y=\min \{y| \ V_{x''-i}\subset \mathcal{T}_{y}; \ i=1,2 \}$. Hence, for all $V_{x''}\subset \mathcal{T}_{y}$; $x+3>x''>x'+1$ we have
  $y''>y'$. Thus, if $V_{x+3}\subset \mathcal{T}_y$ for $y>u_x$, the edges connecting the vertices in $V_{x+2}$ and $V_{x+3}$ would have to cross a line twice.

  If $u_{x}$ was the most recently set in $I_{x'+1}$ such that in the $x'$-th step  case (ii) a) applies, $(x'+1)$-st step has to be again followed by a step
  in which case (iv) a) applies. So, we are done by the same argument.

  Finally, if $u_x$ was the most recently set in the $x'$-th step, in which case (iv) b) applies, $(x'+1)$-st step can be
  a step in which either case (i) or case (ii) b) applies. Hence, we can proceed as in the two previous situations.

  This finishes the proof of the lemma if $d=d'$. \\

If $d>d'$, the lemma follows easily, since for every maximal subtree $T'$ of $T_d'$  rooted at a vertex, which has 0 in-degree in $T'$ and is at distance at most $d-d'$ from the root $r$, the above proof goes through.
\end{proof}

We let $\mathcal{T}=\mathcal{T}(\mathcal{P})$ denote a maximal length tubus of $T_{d'}=T'(d',T(d,\Delta))$ starting at $v$, at distance at most $d-d'-1$ from $r$,
having the vertices ($\not= v$) of its defining  paths colored by elements of a 20-element subset of $\{1,\ldots, c\}$.
We denote by $\mathcal{T}_i$  the  $i$-th maximal tubus polygon   of $\mathcal{T}$.
We prove a property of $\mathcal{T}$ following easily from Lemma~\ref{lemma:tubusPolygon}, which, roughly speaking, says that  $\mathcal{T}$ has to intersect lines in at least $(c-19)$  different classes $\mathcal{L}_{c'}$ between its two consecutive cuts by a line.

Suppose that $\mathcal{P}$ has at least one central path. We say that a line {\em cuts} a tubus polygon if it contains a line segment that cuts the tubus polygon.
\begin{lemma}
\label{lemma:tubusPolygon2}
If there exists a line $l$ that cuts $\mathcal{T}_i$ and $\mathcal{T}_{j}$, $c<i\le j<k$, resp., in a line segment (if $i=j$, we require two such line segments),
then $\bigcup_{i'=i}^{j}\mathcal{T}_{i'}$ intersects lines in at least $(c-19)$ different classes $\mathcal{L}_{c'}$, $1\leq c'\leq c$.
\end{lemma}

\begin{proof}
\bigskip
\begin{figure}[htb]
\centering
\includegraphics[scale=0.5]{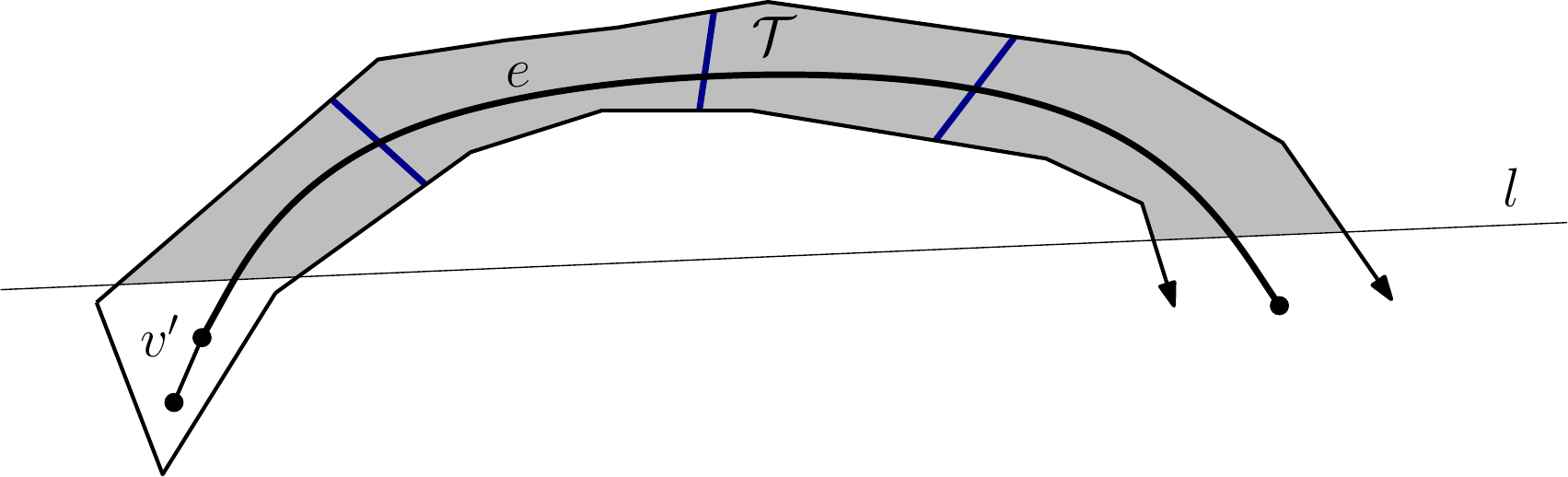}
\caption{The edge $e$ is forced to cross the line $l$ twice.}
\label{fig:3paths}
\end{figure}

We proceed by a contradiction. 
Thus, let $C$ denote the subset of $\{1,\ldots, c\}$ of size 20 such that the lines in $\bigcup_{c'\in C} \mathcal{L}_{c'}$, are not intersected $\bigcup_{i'=i}^{j}\mathcal{T}_{i'}$.

Let $v'\in V_1$ denote a vertex lying on a central path $P$ of $\mathcal{P}$. Note that $V_1\subseteq \mathcal{T}_{i''}$, for $i''< c$.
Let $\mathcal{T}'$ denote a maximum length tubus starting at $v'$, which is defined by paths in the maximal subtree
$T'$ of $T_{d'}$ of depth $d'$ rooted at  $v'$ (having 0 in-degree in $T'$), whose vertices ($\not=v'$) are colored by the elements of $C$.

By Lemma~\ref{lemma:tubusPolygon}, 
both $\mathcal{T}'$ and $\mathcal{T}$ have length of at least  $k$. Thus, as $v'$ is on a central path of
$\mathcal{T}$, we have $\mathcal{T}'\cap \mathcal{T}_{i'} \not= 0$, for $i\le i'\le j+1$. Hence, an edge on a path contained in $\mathcal{T}'$ (see Fig.\ref{fig:3paths}) has to cross $l$ twice (contradiction).
\end{proof}

\begin{figure}[htb]
\centering
\subfigure[]{
  \includegraphics[scale=0.4]{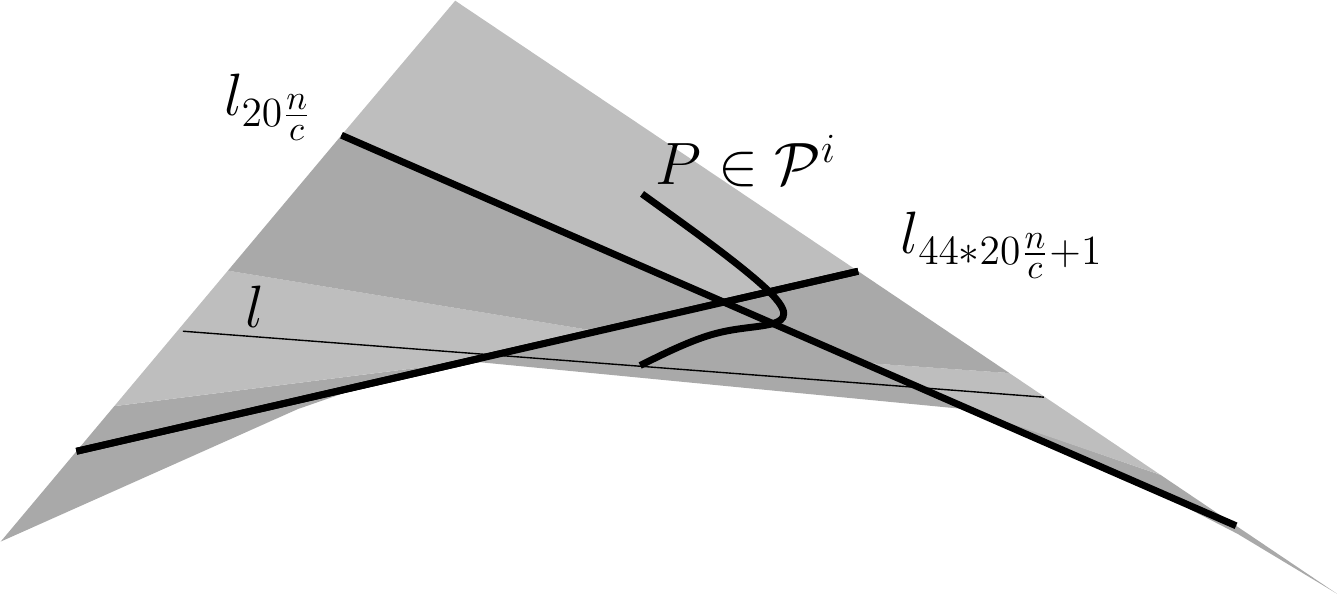}
  \label{fig:bicoloredAreas3}} \hspace{1cm}
   \subfigure[]{
   \includegraphics[scale=0.5]{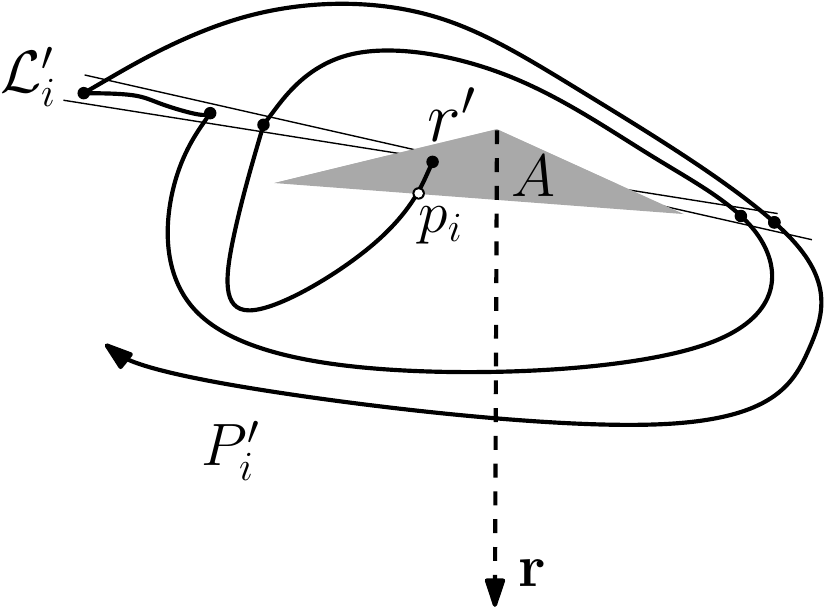}
  \label{fig:bicoloredAreas4}} \hspace{5mm}
   \subfigure[]{
   \includegraphics[scale=0.4]{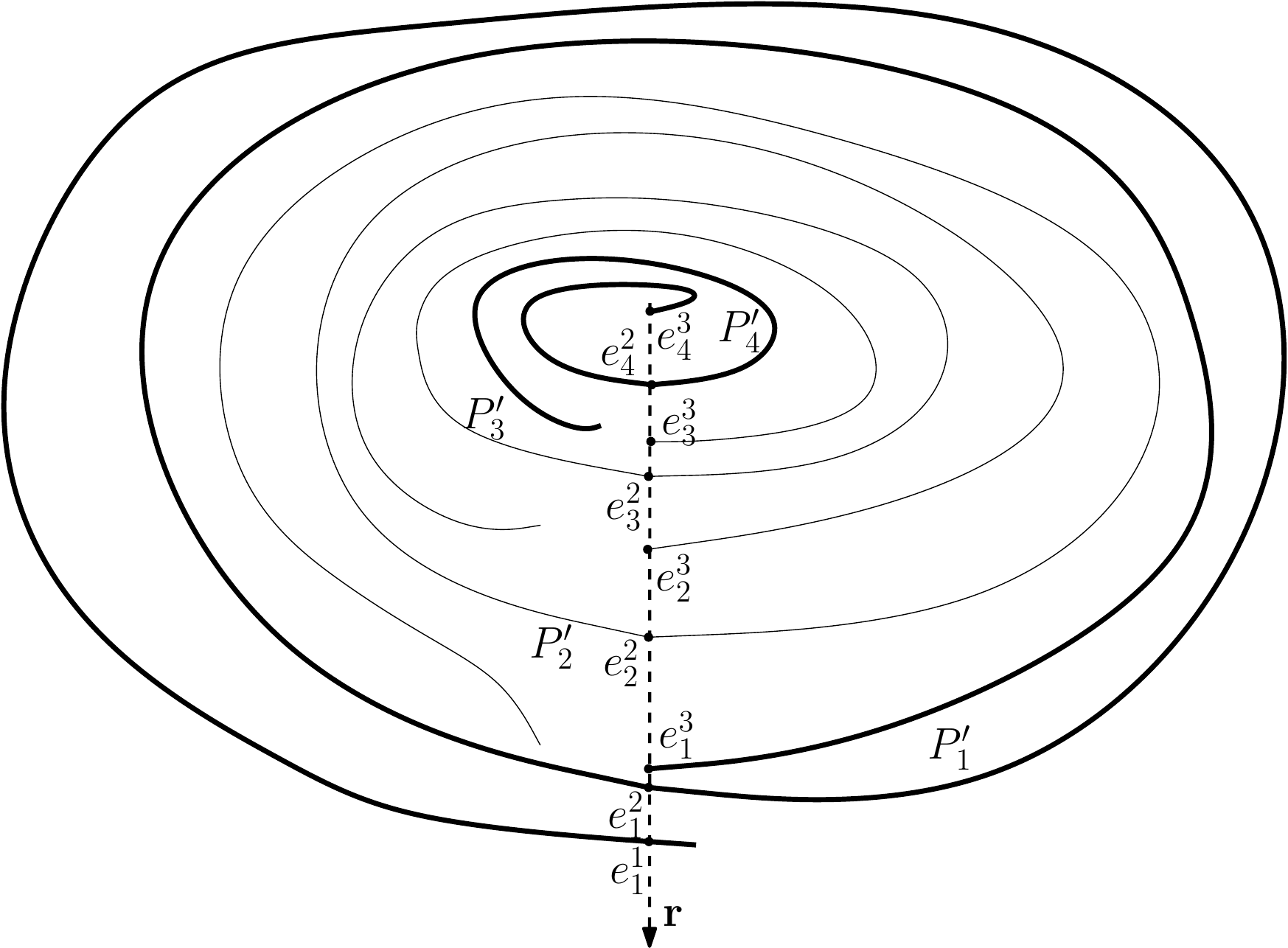}
  \label{fig:bicoloredAreas5}}
  \caption{(a) The union of the regions $\overline{R_{a,b}}$, for which $\frac {c}{45}+1 > a$ or $b > 44\frac {c}{45}$,
bounded by thick lines, (b) $P_i'$ winding around $A$, (c) $P_1$ and $P_4$ cannot meet at $r'$.}
\end{figure}

Before we proceed with the rest of the proof of Theorem~\ref{thm:concurrentLines2} we need to introduce some additional notions.
Let $A$ denote an oriented Jordan arc in the plane and ${\bf r}$ denote a ray (i.e. half-line) in the plane. 
We define the \emph{winding number} of $A$ with respect to ${\bf r}$ as the number of times  we  arrive at ${\bf r}$ from the left side (when looking in the direction of ${\bf r}$) minus the number of times we arrive at ${\bf r}$ from the right side, if we traverse $A$ consistently with its orientation.
We say that $A$ \emph{winds} $c$ times w.r.t. ${\bf r}$ if its winding number w.r.t. ${\bf r}$ is $c$.

Let $S\subseteq  \mathbb{R}^2$ denote a simply connected compact set such that $A\cap S=\emptyset$.
We say that $A$ \emph{winds $c$ times around} $S$ if its winding number w.r.t. to a ray ${\bf r}$ emanating from a point in $S$ is at least $c$.

Using Lemma~\ref{lemma:tubusPolygon2} it is not hard to see that a set of super uniform paths winds around the convex hull of the intersection points  of a big subset of $\mathcal{L}$:

Let $\mathcal{L}_1',\ldots,\mathcal{L}_{45}'$ denote the subsets of $\mathcal{L}$ such  that $\mathcal{L}_i'=\bigcup_{j=(i-1)\frac {c}{45} +1}^{i\frac {c}{45}}\mathcal{L}_j$.
Let $d'=d'(k)+1< d,\Delta=\Delta_0(k,c,3)$, and $c=20* 45$ ($d'(k)$ and $\Delta_0(k,c,3)$ as in Lemma~\ref{lemma:tubusPolygon}).  Let $\mathcal{T}^3(\mathcal{P}^3),\ldots, \mathcal{T}^{43}(\mathcal{P}^{43})$ denote the tubuses of $T_{d'}=T'(d',T(d,\Delta))$ of length $k$ with 3 central paths, for $k$ (specified later) sufficiently big with respect to $c$, whose defining paths start at an arbitrary child  $r'\in T_{d'}$ of the root $r$, so
that the paths defining $\mathcal{T}^i$ have the vertices ($\not=r'$) mapped by $\iota$ to $\mathcal{L}_i'$.
Let $A$ denote the  union of  regions $\overline{R_{a,b}}$; $2\frac{c}{45}+1 \leq a \leq b \leq 43\frac{c}{45}$.

\begin{lemma}
\label{lemma:noA}
Let $k>10c^2+c$.
For all $i,j:c<j<k-10c^2$, $3\leq i \leq 43$, $\mathcal{T}_j^i$ is not contained in   $A$.
\end{lemma}

\begin{proof}
For the sake of contradiction let us choose a minimal $j:c<j<k-10c^2$, such that $\mathcal{T}_j^i\subseteq \overline{R_{a,b}}; \ 2\frac{c}{45}+1 \leq a \leq b \leq 43\frac{c}{45}$. Suppose that $\mathcal{T}_j^i$ is above the line $l$ containing the door between $\mathcal{T}_j^i$ and $\mathcal{T}_{j-1}^i$ (the below-case is treated analogously).
Thus, $l$ cuts $\mathcal{T}_j^i$. If for all $\mathcal{T}_{j'}^i$, $j':j<j'<j+5c^2$, $\mathcal{T}_{j'}^i\subseteq \overline{R_{a,b}}; \ \frac{c}{45}+1 \leq a \leq b \leq 44\frac{c}{45}$,
we are in contradiction with Lemma~\ref{lemma:tubusPolygon2} (by Pigeon Hole principle).

Hence, we have $\mathcal{T}_{j'}^i\subseteq \overline{R_{a',b'}}$ such that $\frac {c}{45}+1 > a'$ or $b' > 44\frac {c}{45}$, for some $j':k-5c^2>j'>j$. Let us choose $j'$ as small as possible.
Observe that $\mathcal{T}_{j'}^i$ is above $l$, by Lemma~\ref{lemma:tubusPolygon2}. Let's say $\frac {c}{45}+1 > a'$.
Ref. to Fig.~\ref{fig:bicoloredAreas3}.
By Lemma~\ref{lemma:tubusPolygon2}, applied to the line $l_{20\frac nc}$, 
for every $j''>j'$,
 $\mathcal{T}_{j''}^i\subseteq \overline{R_{a'(j''),b'(j'')}}$ such that $\frac {c}{45}+1 > a'(j'')$, unless
   for some $j''>j'$, $\mathcal{T}_{j''}^i\subseteq \overline{R_{a'(j''),b'(j'')}}$
  so that $44\frac{c}{45}< b'(j'')$ and $\frac {c}{45}+1 > a'(j'')$, and  for some $j''>j'$, $\mathcal{T}_{j''}^i\subseteq \overline{R_{a'(j''),b'(j'')}}$
  so that $2\frac{c}{45}+1> b'(j'')$ and $\frac {c}{45}+1 > a'(j'')$. However, in what follows we show that this is not possible without
  $l$ or $l_{44*20\frac cn+1}$ cutting tubus polygons so that Lemma~\ref{lemma:tubusPolygon2} is violated, by the choice of $j'$.

   To this end we distinguish two cases, either we first
  visit a region $\overline{R_{a',b'}}$ such that $2\frac{c}{45}+1> b'$ and $\frac {c}{45}+1 > a'$, or
  we first visit a region $\overline{R_{a',b'}}$ such that $44\frac {c}{45}< b'$ and $\frac {c}{45}+1 > a'$.
  In the latter, Lemma~\ref{lemma:tubusPolygon2} is violated by cutting with the line $l_{44*20\frac cn+1}$, in the former with the line $l$.
  It follows that for no $j''>j'$, $\mathcal{T}_{j''}^i\subseteq \overline{R_{a'(j''),b'(j'')}}$, where $44\frac{c}{45}< b'(j'')$ and $\frac{c}{45}+1 > a'(j'')$,
  or for no $j''>j'$, $\mathcal{T}_{j''}^i\subseteq \overline{R_{a'(j''),b'(j'')}}$, where $2\frac{c}{45}+1> b'(j'')$ and $\frac{c}{45}+1 > a'(j'')$.

  Thus, $\mathcal{T}_{j''}^i$  for all $j''>j'$ does not intersect a line in $\mathcal{L}'_2$, or  $\mathcal{T}_{j''}^i$  for all $j''>j'$ does not intersect a line in $\mathcal{L}'_{45}$.
  On the other hand, by Pigeon hole principle and the condition $j'<k-5c^2$ we violate Lemma~\ref{lemma:tubusPolygon2} (contradiction).
\end{proof}

Ref. to Fig.~\ref{fig:bicoloredAreas4}.
We denote by ${\bf r}$ a vertical ray emanating from the point in $A$ that is not below any line in $\cup_{i=3}^{43}\mathcal{L}_i'$   heading downward.
Observe that the lines in $\mathcal{L}_i'$ participate only in 20 regions of the form $\overline{R_{a',b'}}$, such that  $\frac {c}{45}+1 > a'$
(resp. $b' > 44\frac {c}{43}$), each of which has at most 5 sides through which the paths in $\mathcal{P}^i$ can enter it. Thus, by Lemma~\ref{lemma:tubusPolygon2} and
Lemma~\ref{lemma:noA} we increase (resp. decrease) the winding
number of the paths in $\mathcal{P}^i$ w.r.t. ${\bf r}$ at least once per $2(5*20+c)$ tubus polygons except for the first
$c$ or last $10c^2$ tubus polygons of a tubus.
Hence, by letting $\Delta$ and $k'$ grow we can find for every $3\leq i\leq 43$ the set of paths $\mathcal{P}^i$ in $T_{d'}$ such that the paths in each set $\mathcal{P}^i$ have internal vertices mapped by $\iota$ to $\mathcal{L}_i'$; emanate from a child $r'$ of the root; are internally disjoint;  and their subpaths wind many times around $A$.

\bigskip

In the rest of the proof we proceed in three stages.
First, we select three subpaths of paths from three different sets $\mathcal{P}^i$ that winds many times around $A$ in an interleaved fashion.
Second, having for each child of the root in $T_{d'}$ three ``good'' paths we pick three children $r_1,r_2$ and $r_3$ of the root, whose three ``good'' paths are consistent in some sense, as specified later. Finally, among nine ``good'' paths emanating from $r_1,r_2$ and $r_3$ we pick three paths with edges, on which Lemma~\ref{lemma:unstretch} can be applied. We proceed with the first stage.

\paragraph{Stage 1}
If $k>12(5*20+c)+
10c^2+c$, we can pick 41 subpaths  $P_3,\ldots, P_{43}$  of the paths from $\mathcal{P}^3,\ldots, \mathcal{P}^{43}$, resp., starting at $r'$ so that the following holds: $P_i$, $3\leq i \leq 43$, has the winding number w.r.t. ${\bf r}$ at least 6 or at most -6.
 Let $P_i'$, $3\leq i \leq 43$, denote a subpath of $P_i$ in the topological sense completely disjoint from the interior of $A$ starting at a point $p_i$, which belongs to the boundary of $A$ or coincides with $r'$. By Lemma~\ref{lemma:noA}, we can select $P_i'$ so that it has the winding number w.r.t. ${\bf r}$ at least 5 or at most -5. (The initial pieces of the paths $P_i$ traversing the first $c$ tubus polygons defined by $\mathcal{P}^i$ account for the difference by at most 1.)

W.l.o.g. 21 paths among $P_2',\ldots P_{42}'$ winds w.r.t. ${\bf r}$ at least 5 times.
Note that no subpath in the topological sense of such a path has the winding number w.r.t ${\bf r}$ less than -1, as otherwise we would contradict Lemma~\ref{lemma:tubusPolygon2}. This property is crucial for the rest of the proof.
 We show  that 3 of these 21 paths, let us relabel them as $P_1',P_2'$ and $P_3'$, wind around $A$ in an interleaved fashion:

  Let $e_i^j$ denote the intersection point of $P_i'$ with ${\bf r}$, such that the subpath $P_i''$ of $P_i'$ with the same starting point $p_i$
and ending  at $e_i^j$ has the winding number $j$ w.r.t. ${\bf r}$. Let us choose $e_i^j$-s so that $P_i''$-s are as short as possible.

\begin{proposition}
\label{prop:interleave}
We can choose $P_1',P_2'$ and $P_3'$ so that the points  $e_1^2, e_2^2, e_3^2,e_1^3, e_2^3, \ldots, e_3^5$ appear (possibly after relabeling the paths) along ${\bf r}$ in that order.
\end{proposition}

\begin{proof}
Observe that all the points $e_i^j$, for $j>2$ and $i$ fixed,  are in one connected component of ${\bf r}\setminus \{e_i^2\}$.
Really, otherwise we obtain a tubus violating Lemma~\ref{lemma:tubusPolygon2} (see Fig.~\ref{fig:bicoloredAreas4}).
Thus, we can choose 11 paths $P_1',\ldots ,P_{11}'$ among our 21 paths winding in the same sense around $A$ such that for every  $i\in\{1,\ldots,21 \}$, all $e_i^j$, $j>2$, are contained either in the bounded or in the unbounded connected component of ${\bf r}\setminus \{e_i^2\}$.  Let us assume that the former happens (the latter is treated analogously).
 Let us define the poset $(\{P_1',\ldots ,P_{11}'\},<)$, where $P_i'<P_j'$, if $e_i^2, e_i^3, e_j^2,e_j^3$ appear along ${\bf r}$
(from bottom to top) in that order.

It is left to show that among these 11 paths we can choose $P_1',P_2'$ and $P_3'$ such that \\ $e_1^2, e_2^2, e_3^2,e_1^3, e_2^3, e_3^3$ appear along ${\bf r}$ from
the bottom to the top in that order.
Indeed, once we show that, it must be that $e_1^2, e_2^2, e_3^2,e_1^3, e_2^3, \ldots, e_3^5$ appear along ${\bf r}$ (from bottom to top) in that order.
If we cannot choose such three  paths, by Dilworth Theorem, we can choose four paths $P_1',P_2', P_3'$ and $P_4'$ so that their intersection points appear along ${\bf r}$ in the following order: $e_1^2, e_1^3,  e_2^2, e_2^3, e_3^2, e_3^3, e_4^2, e_4^3$  (see Fig.~\ref{fig:bicoloredAreas5}), since $e_i^2, e_j^2, e_j^3,e_i^3$ cannot appear in this order for any pair $i,j$.
Now, the paths $P_1$ and $P_4$ cannot meet in $r'$(contradiction). Indeed, by a bit tedious but easy argument using Jordan Curve Theorem, which we omit in this version,  $P_4$ has to wind w.r.t. ${\bf r}$ more than 2 times on a subpath ending at $e_4^2$, thereby contradicting the choice of $e_4^2$, or $P_1$ has to wind w.r.t. ${\bf r}$ less than -1 times on a subpath (contradiction).
Hence, we obtained three paths $P_1,P_2$ and $P_3$ winding around $A$ in the interleaved fashion.
\end{proof}

Let us denote the paths  $P_1',P_2'$ and $P_3'$, respectively, from the previous proposition by $P_1^{r'}, P_2^{r'}$ and $P_3^{r'}$, respectively.
Let us denote the corresponding paths $P_{i_1},P_{i_2}$ and $P_{i_3}$, respectively, containing $P_1',P_2'$ and $P_3'$, respectively,  by $Q_{1}^{r'}, Q_{2}^{r'}$ and $Q_{3}^{r'}$, respectively.
Analogously, in what follows we refer to $e_i^j$ as to $e_i^{j,r'}$.

\paragraph{Stage 2} At this point we have for each child $r'$ of the root in $\mathcal{T}_{d'}$ three good paths $Q_{1}^{r'}, Q_{2}^{r'}$ and $Q_{3}^{r'}$.
In the following, we select three children of the root $r_1,r_2$ and $r_3$, such that for each of them the three paths $Q_{1}^{r_i}, Q_{2}^{r_i}$ and $Q_{3}^{r_i}$ wind ``consistently'' around $A$ and have the vertices mapped by $\iota$ to the same three sets $\mathcal{L}_{j_1}', \mathcal{L}_{j_2}'$ and $\mathcal{L}_{j_3}'$, respectively. Here, by ``consistently'' we mean that they wind around $A$
in the same sense and that the points $e_{1}^{2,r_3},e_{{2}}^{2,r_3},e_{{3}}^{2,r_3}, e_{{1}}^{2,r_2}, e_{{2}}^{2,r_2}, e_{{3}}^{2,r_2}, e_{{1}}^{2,r_1},  e_{{2}}^{2,r_1}, e_{{3}}^{2,r_1},\ldots,
e_{{1}}^{5,r_2},e_{{2}}^{5,r_2},\\ e_{{3}}^{5,r_2}, e_{{1}}^{5,r_1}, e_{{2}}^{5,r_1},e_{{3}}^{5,r_1}$ appear along ${\bf r}$ in this order.

By the pigeon hole principle paths $P_1^{r'}, P_2^{r'}$ and $P_3^{r'}$ wind around $A$ in the same direction and the corresponding intersection points $e_i^{j,r'}$, $j>2$,
lie in the bounded (resp. unbounded) component of ${\bf r}\setminus \{e_i^{2,r'}\}$ for at least $\frac 14$ of the children $r'$ of $r$.
Let us denote such a set of children of the root by $V_r$ and let us assume that $P_i^{r'}$, $r'\in V_r$, winds w.r.t. ${\bf r}$ at least 5 times such that
 $e_i^{j,r'}$, $j>2$, lie in the bounded component of ${\bf r}\setminus \{e_i^{2,r'}\}$  (the other case is treated analogously).
  Let $\mathcal{P}$ denote the set of the paths $P_i^{r'}$, $r'\in V_r$, $i=1,2,3$.

The following proposition follows easily by Dilworth theorem similarly as Proposition~\ref{prop:interleave}.

\begin{proposition}
\label{prop:Vrprime}
We can partition $V_r$ into at most five parts so that for every $r'$ and $ r''$ belonging to the same part $e_{i'}^{2, r'}, e_{i''}^{2,r''}, e_{i'}^{3, r'}, e_{i''}^{3, r''}$ (resp. $e_{i''}^{2, r''}, e_{i'}^{2,r'}, e_{i''}^{3, r''}, e_{i'}^{3, r'}$), for $i',i''\in\{1,2,3\}$,  appear along ${\bf r}$ in that order.
\end{proposition}

Let $V_r'\subseteq V_r$ denote the biggest part of the partition from the last proposition.
Let $\mathcal{P}'\subseteq \mathcal{P}$ correspond to $V_r'$, i.e. $P_i^{r'}\in \mathcal{P}'$, for $r'\in V_r'$.
Note that $V_r'$ contains at least $\frac 1{20}$ of the children of $r$.
The next proposition rules out the possibility of two pairs of paths $P_{i_1}^{r'},P_{i_3}^{r'}$ and  $P_{i_2}^{r''},P_{i_4}^{r''}$,
$r',r''\in V_r'$; $i_1,i_2,i_3,i_4\in \{1,2,3\}$, $i_1\not=i_3$ and $i_2\not=i_4$ to ``interleave'' in some sense:

\begin{proposition}
\label{prop:goodOrder}
The points $e_{i_1}^{2,r'}, e_{i_2}^{2,r''}, e_{i_3}^{2,r'}, e_{i_4}^{2,r''}$ cannot appear along ${\bf r}$ in that order for any $r',r''\in V_r'$;
$i_1,i_2,i_3,i_4\in \{1,2,3\}$.
\end{proposition}

\begin{figure}[htp]
\centering
\subfigure[]{\includegraphics[scale=0.4]{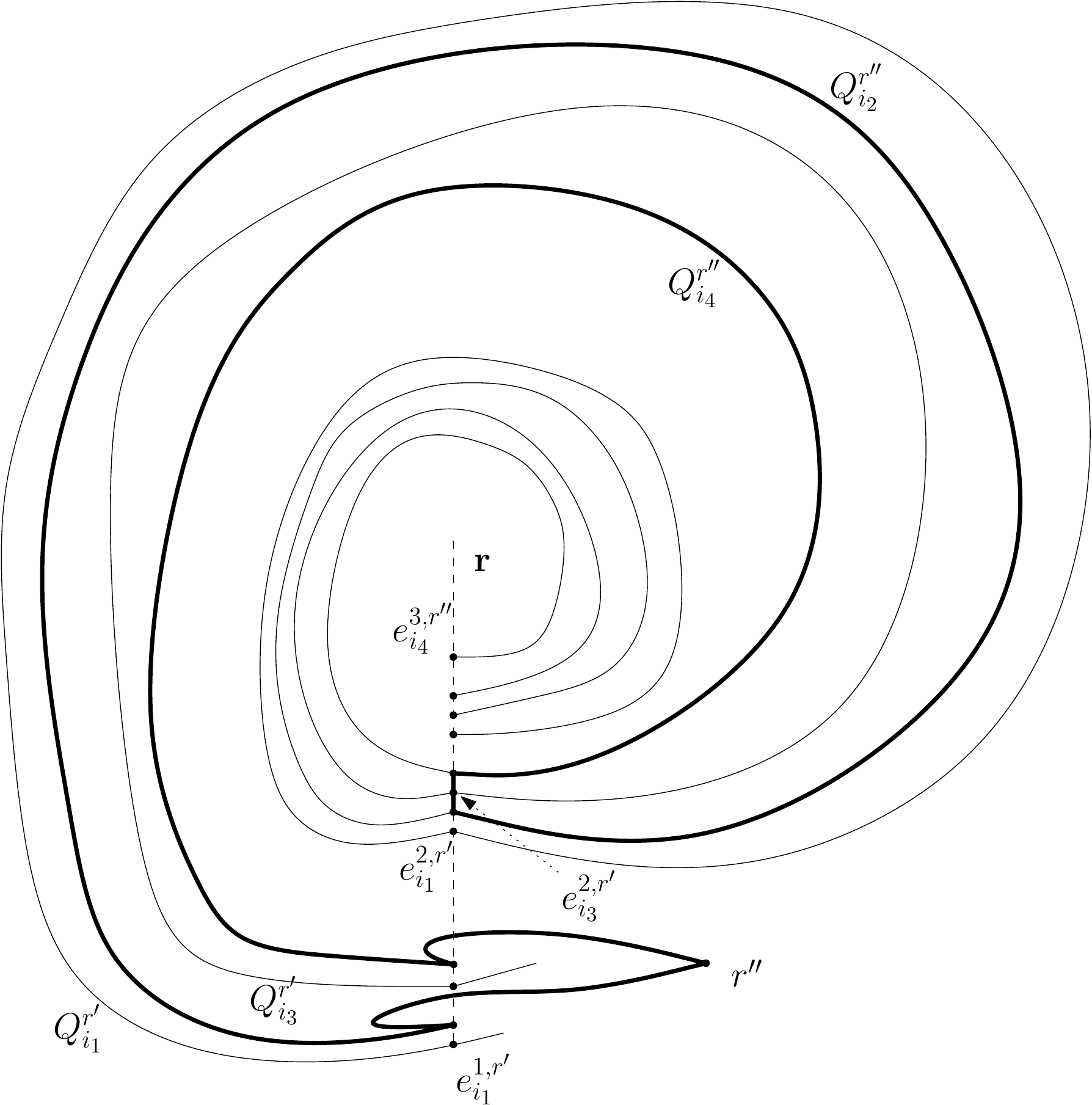}
\label{fig:prop4}}
\subfigure[]{\includegraphics[scale=0.4]{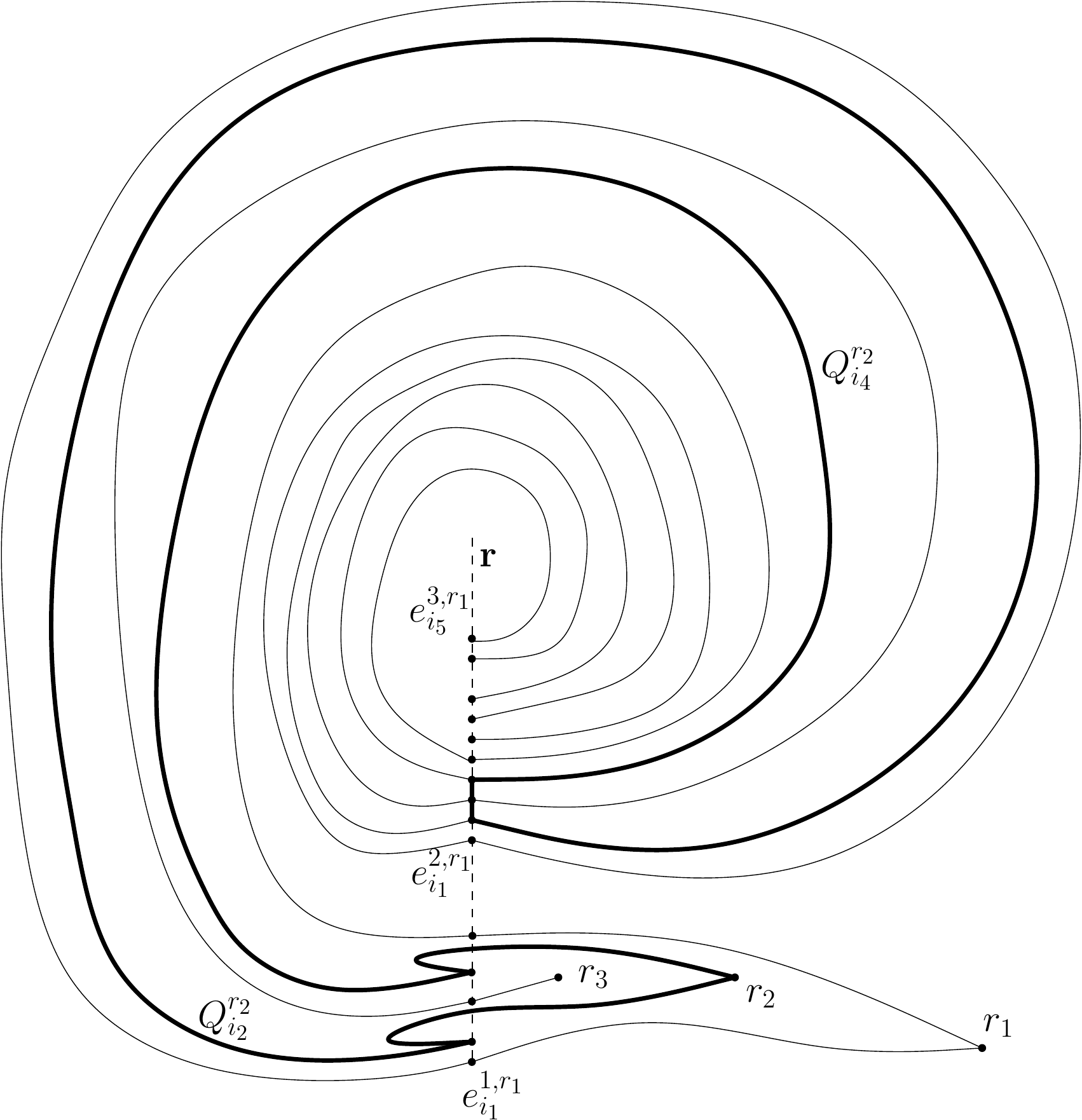}
\label{fig:prop5}}
\caption{(a) Illustration for Proposition~\ref{prop:goodOrder}, (b) Illustration for Proposition~\ref{prop:goodOrder2}}
\end{figure}

\begin{proof}
Ref. to Fig.~\ref{fig:prop4}.
If we have $r',r'',i_1,i_2,i_3$ and $i_4$ contradicting the claim, $Q_{i_1}^{r'}$ cannot meet $Q_{i_3}^{r'}$ in $r'$, since
$Q_{i_3}^{r'}$ is up to $e_{i_3}^{2,r'}$ completely contained in a compact region bounded by Jordan arc consisting of the segment $e_{i_2}^{2,r''}e_{i_4}^{2,r''}$ and two curves starting at $r''$ corresponding  to subpaths of $Q_{i_2}^{r''}$ and $Q_{i_4}^{r''}$, while $Q_{i_1}^{r'}$
is up to $e_{i_1}^{2,r'}$  completely outside of this region.
\end{proof}

Moreover, at most two pairs of paths $P_{i_1}^{r'},P_{i_3}^{r'}$ and  $P_{i_2}^{r''},P_{i_4}^{r''}$ can be ``nested'':

\begin{proposition}
\label{prop:goodOrder2}
The points $e_{i_1}^{2,r_1}, e_{i_2}^{2,r_2}, e_{i_3}^{2,r_3} , e_{i_4}^{2,r_2}, e_{i_5}^{2,r_1}$ cannot appear along ${\bf r}$ in that order for any $r_1,r_2,r_3\in V_r'$;
$i_1,i_2,i_3,i_4,i_5\in \{1,2,3\}$.
\end{proposition}

\begin{proof}
Ref. to Fig.~\ref{fig:prop5}.
Having $r_1,r_2,r_3,i_1,i_2,i_3,i_4$ and $i_5$ contradicting the claim the root $r$ has to be contained simultaneously outside (because of $r_1$) and inside (because of $r_3$) of the compact region bounded by a Jordan arc consisting of the segment $e_{i_2}^{2,r_2}e_{i_4}^{2,r_2}$ and two curves starting at $r_2$ corresponding to subpaths of $Q_{i_2}^{r_2}$ and $Q_{i_4}^{r_2}$ (contradiction).
\end{proof}

Finally, by Proposition~\ref{prop:goodOrder} we can select five vertices $r_1,\ldots, r_5\in V_r'$ (provided $|V_r'|$ is at least
$4{41 \choose 3}+1$, which is satisfied due to the choice of parameter $\Delta$) such that the paths $P^{r_j}_{1},P^{r_j}_{2}, P^{r_j}_{3}$, respectively, have
 the vertices mapped by $\iota$ to the same classes $\mathcal{L}_{j_1}', \mathcal{L}_{j_2}', \mathcal{L}_{j_3}'$, respectively, and no two pairs of paths of the form  $P^{r_j}_{i_1},P^{r_j}_{i_2}$ ``interleave'', i.e. $e_{i_1}^{2,r_1}, e_{i_2}^{2,r_2}, e_{i_3}^{2,r_1}, e_{i_4}^{2,r_2}$
 appear along ${\bf r}$ in that order.
It can still happen that two pairs of paths of the form $P^{r_j}_{i_1},P^{r_j}_{i_2}$ are nested, though,
i.e. $e_{i_1}^{2,r_1}, e_{i_3}^{2,r_2}, e_{i_4}^{2,r_2}, e_{i_2}^{2,r_1}$, $i_3\not=i_4$, appear along ${\bf r}$ in that order.
By Proposition~\ref{prop:goodOrder} we can define the poset $(\{r_1,\ldots, r_5\},<)$ such
 that $r_i<r_j$ if $e_{i_1}^{2,r_i}, e_{i_3}^{2,r_j}, e_{i_4}^{2,r_j}, e_{i_2}^{2,r_i}$, for some $i_1,i_2,i_3,i_4\in \{1,2,3\}$; $i_3\not=i_4$, appear along ${\bf r}$ in that order.
By Proposition~\ref{prop:goodOrder2} our poset does not have a chain of size 3. Hence, by Dilworth theorem we can pick (w.l.o.g) $r_1,r_2$ and $r_3$ such that $$e_{1}^{2,r_3},e_{{2}}^{2,r_3},e_{{3}}^{2,r_3}, e_{{1}}^{2,r_2}, e_{{2}}^{2,r_2}, e_{{3}}^{2,r_2}, e_{{1}}^{2,r_1},  e_{{2}}^{2,r_1}, e_{{3}}^{2,r_1},\ldots,
e_{{1}}^{5,r_2},e_{{2}}^{5,r_2},e_{{3}}^{5,r_2},e_{{1}}^{5,r_1},e_{{2}}^{5,r_1},e_{{3}}^{5,r_1}$$  appear in that order along ${\bf r}$.

\paragraph{Stage 3} Let $m$ denote a 3 by 3 matrix over the set $\{1,2,3\}$ not containing two equal elements in the same row or column such that
 the vertices ($\not=r_j$) of the paths in the set
 $\mathcal{S}_i= \{ P_{m(i,j)}^{r_j}\ |j=1,2,3 \}$   are for each $i=1,2,3$  mapped by
$\iota$ to the same class $\mathcal{L}_{j_i}'$. Let $j_1,j_2$ and $j_3$ denote the indices of the corresponding classes $\mathcal{L}_{j_1}', \mathcal{L}_{j_2}'$ and $\mathcal{L}_{j_3}'$, so that the slopes of the lines in $\mathcal{L}_{j_{x+1}}'$ are bigger
than in $\mathcal{L}_{j_{x}}'$ for $x=1,2$.

\begin{figure}[htp]

  \centering
   \subfigure[]{\includegraphics[scale=0.3]{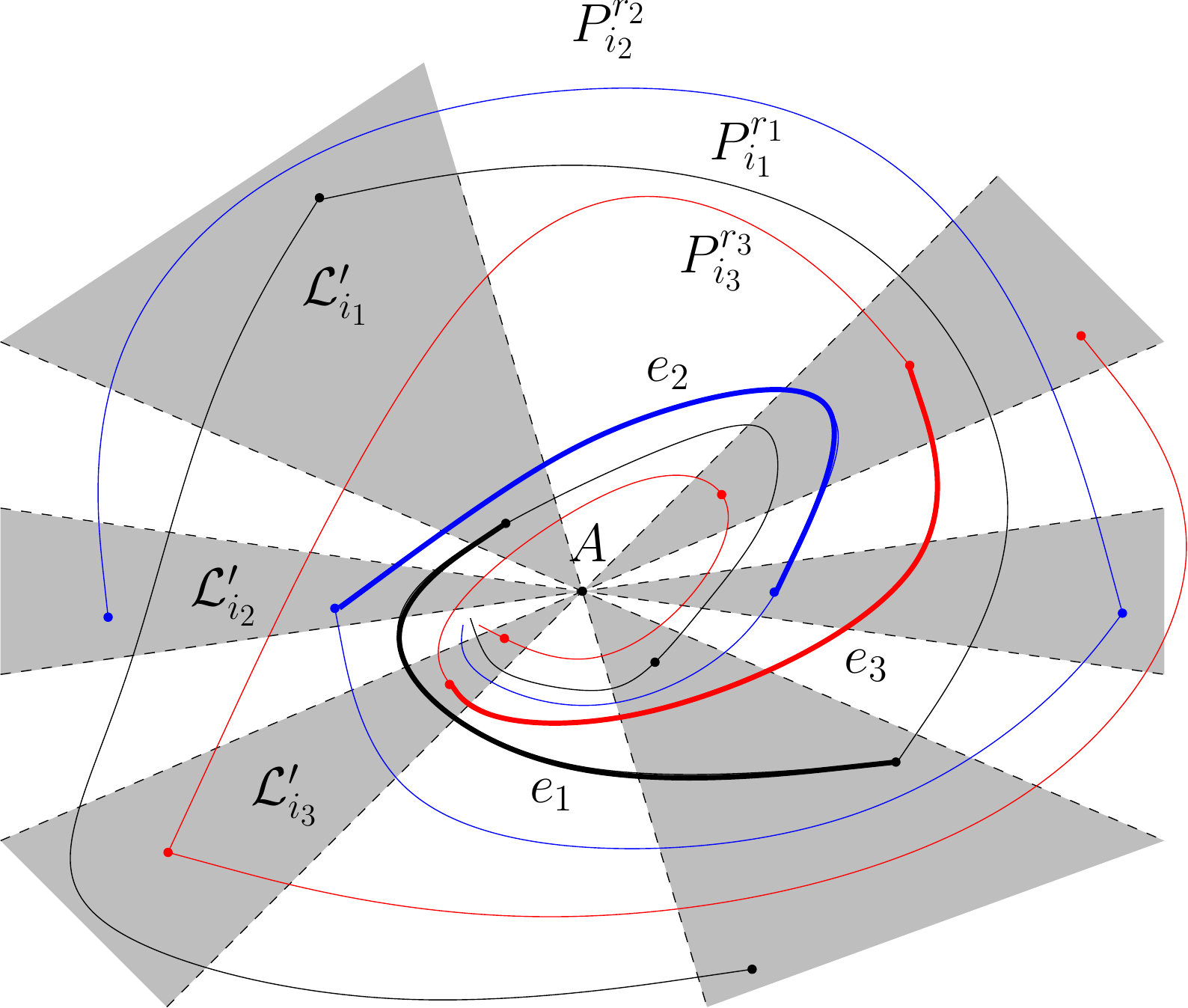}
	  \label{fig:deadlock1e} \hspace{5mm}
	}
\subfigure[]{\includegraphics[scale=0.3]{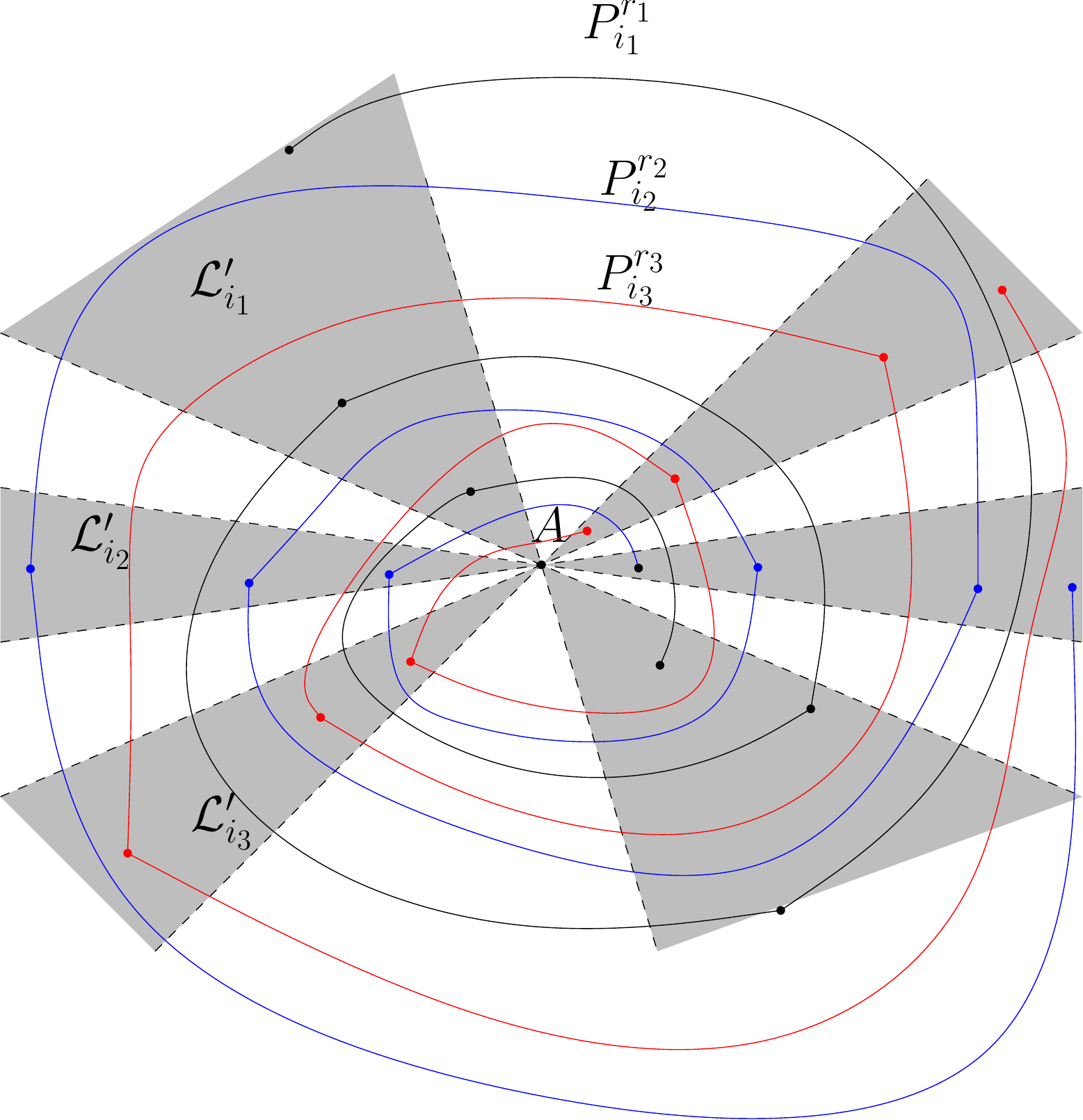}
	  \label{fig:deadlock1e2} \hspace{5mm}
	}
   \subfigure[]{\includegraphics[scale=0.3]{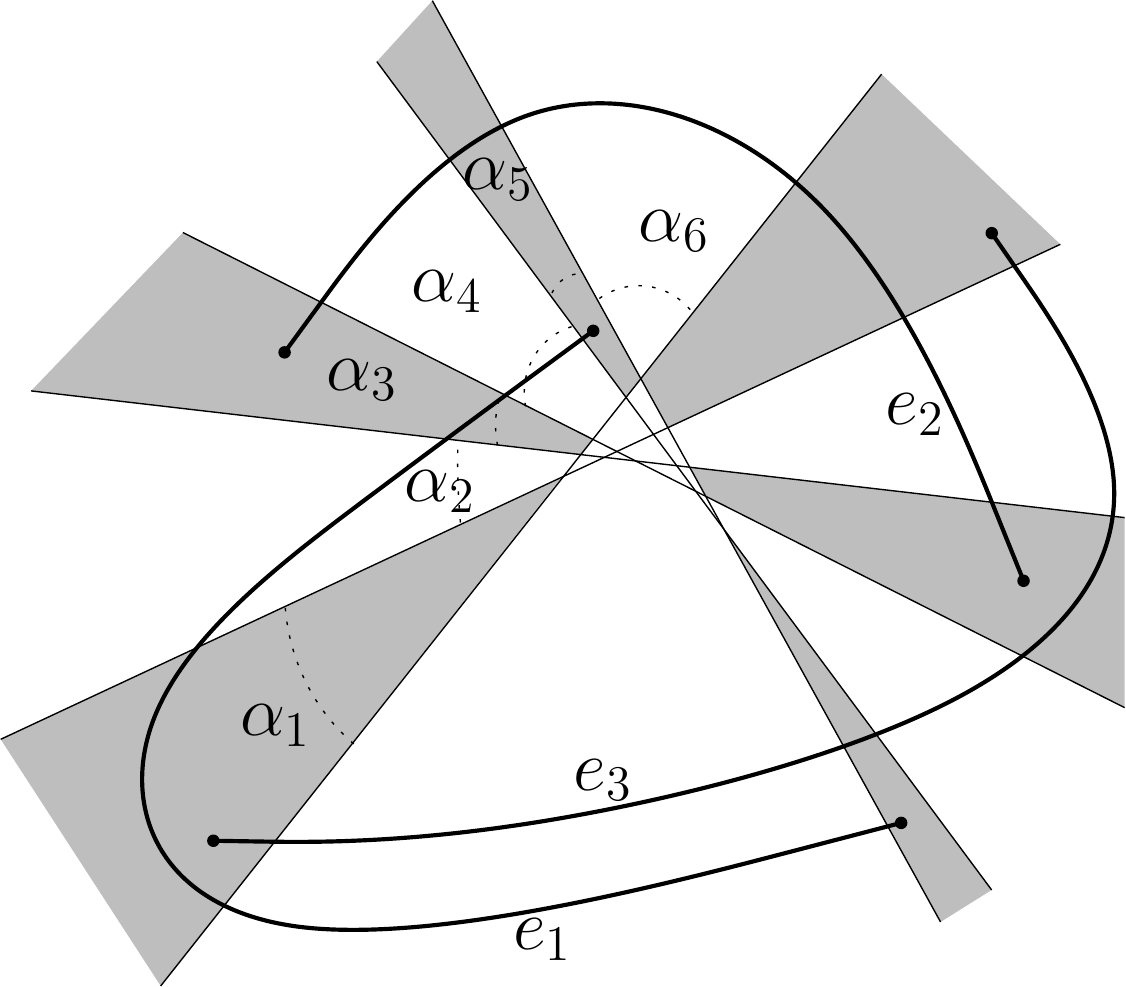}
	  \label{fig:deadlock1eee}}
   \caption{(a) Paths winding around $A$ (due to a better readability the edges $e_1,e_2$ and $e_3$ are drawn as arcs, even though they are, of course, the straight-line segments and $A$ is represented by a point); (b) The situation we want to avoid; (c) The edges $e_1,e_2$ and $e_3$ connecting the grey regions.}

\end{figure}

We pick three paths $P_{i_1}^{r_1}$, $P_{i_2}^{r_2}$ and $P_{i_3}^{r_3}$, so that the vertices from $P_{i_1}^{r_1}, P_{i_2}^{r_2}, P_{i_3}^{r_3}$, respectively,
 are mapped by $\iota$ to
$\mathcal{L}_{j_{1}}', \mathcal{L}_{j_{2}}', \mathcal{L}_{j_{3}}'$, respectively, (see Fig.~\ref{fig:deadlock1e}).
Note that $e^{2,r_3}_{i_3}, e^{2,r_2}_{i_2}, e^{2,r_1}_{i_1}$ appear along ${\bf r}$ from the bottom to the top in that order, which is consistent with the clockwise order of the lines in $\bigcup_{i}\mathcal{L}_{j_i}'$, in which the lines in $\mathcal{L}_{j_{3}}'$ are followed by the lines in $\mathcal{L}_{j_{2}}'$ that are followed by the lines in $\mathcal{L}_{j_{1}}'$.
Notice that we really needed nine paths $P_{i_{j}}^{r_{j'}}$ in order to avoid the situation from
Fig.~\ref{fig:deadlock1e2}, i.e. $e^{2,r_1}_{i_1}, e^{2,r_2}_{i_2}, e^{2,r_3}_{i_3}$ appear along ${\bf r}$ from the bottom to the top in that order, which is not useful for us as we will see later.

Let us define a relation $<$ (not necessarily antisymmetric) on the edges of $T$ as follows: $e<f$ if there exists a ray emanating from $A$ intersecting $e$ before $f$.
Now it is easy to see that  we can pick three edges $e_1,e_2$ and $e_3$  belonging to $P_{i_1}^{r_1}$, $P_{i_2}^{r_2}$ and $P_{i_3}^{r_3}$, respectively, such that $e_1>e_3>e_2>e_1$ (see Fig.~\ref{fig:deadlock1e}).

Finally, we are in a position to apply Lemma~\ref{lemma:unstretch} on $e_1,e_2$ and $e_3$, since they clearly satisfies the hypothesis of the lemma.
Thus, we can never draw $e_1,e_2$ and $e_3$ as straight-line segments. Hence, there exists no straight-line embedding of $T_{d'}$ respecting $\iota$. This in turn implies that no matter how we extend $\iota$ to $T$ we cannot embed $T$ while respecting $\iota$ and that concludes the proof. \end{proof}

\section{Concluding remarks}

In the paper we negatively answered the question asking whether for all $n$ there exists a set of lines of size $n$ in the plane which is  universal
for  trees. In fact, we proved that for all $n>n_0$ no such a set of lines exists. We tried neither to optimize the value of $n_0$, nor to
estimate it from below by closely analyzing our proof, as the obtained value $n_0$  would be probably very far from being optimal.
Therefore, it would be interesting to have a (possibly much simpler) proof of our result, for which the threshold of $n_0$ would be closer to its smallest possible value.

Since an embedding in our setting is not possible in general, one could look at crossing numbers of straight-line embeddings respecting a given mapping.
As another  direction for future research we propose  to study an analogous problem, in which we  replace lines  by curves defined by polynomials up to a degree $d$, for some constant $d>1$.

\section{Acknowledgement}

We would like to thank Michael Kaufmann for supporting our collaboration and a visit of the first author in T\"ubingen, during which a crucial part of this work was done. Furthermore, we thank to G\'{e}za T\'{o}th and anonymous referees for comments that helped to improve the presentation 
of the result.
\bibliographystyle{plain}
\bibliography{pointsOnLines}

\end{document}